\documentclass[acmtog]{acmart} 
\AtBeginDocument{%
  }

\setcopyright{acmlicensed}
\copyrightyear{}
\acmYear{}
\acmDOI{XXXXXXX.XXXXXXX}

\acmJournal{TOG}
\acmVolume{0}
\acmNumber{0}
\acmArticle{000}
\acmMonth{0}

\usepackage{subcaption}
\usepackage{cleveref}
\usepackage{bm}
\usepackage{amsmath}

\usepackage{printlen}
\usepackage{placeins}

\begin{document}

\title{Computational design of personalized drugs via robust optimization under uncertainty}


\author{Rabia Altunay}
\affiliation{%
  \institution{LUT University}
  \city{Lappeenranta}
  \country{Finland}}
\affiliation{%
 \institution{Turku University of Applied Sciences}
 \city{Turku}
 \country{Finland}}
\email{rabia.altunay@lut.fi}

\author{Jarkko Suuronen}
\affiliation{%
  \institution{LUT University}
  \city{Lappeenranta}
  \country{Finland}}
  \email{jarkko.suuronen@lut.fi}

\author{Eero Immonen}
\affiliation{%
 \institution{Turku University of Applied Sciences}
 \city{Turku}
 \country{Finland}}
 \email{eero.immonen@turkuamk.fi}

\author{Lassi Roininen}
\affiliation{%
  \institution{LUT University}
  \city{Lappeenranta}
  \country{Finland}}
\email{lassi.roininen@lut.fi}

\author{Jari Hämäläinen}
\affiliation{%
  \institution{LUT University}
  \city{Lahti}
  \country{Finland}
}\email{jari.hamalainen@lut.fi}

\renewcommand{\shortauthors}{Altunay et al.}

\begin{abstract}

Effective disease treatment often requires precise control of the release of the active pharmaceutical ingredient (API). In this work, we present a computational inverse design approach to determine the optimal drug composition that yields a target release profile. We assume that the drug release is governed by the Noyes-Whitney model, meaning that dissolution occurs at the surface of the drug. Our inverse design method is based on topology optimization. The method optimizes the drug composition based on the target release profile, considering the drug material parameters and the shape of the final drug. Our method is non-parametric and applicable to arbitrary drug shapes. The inverse design method is complemented by robust topology optimization, which accounts for the random drug material parameters. We use the stochastic reduced-order method (SROM) to propagate the uncertainty in the dissolution model. Unlike Monte Carlo methods, SROM requires fewer samples and improves computational performance. We apply our method to designing drugs with several target release profiles. The numerical results indicate that the release profiles of the designed drugs closely resemble the target profiles. The SROM-based drug designs exhibit less uncertainty in their release profiles, suggesting that our method is a convincing approach for uncertainty-aware drug design.

\end{abstract}


\received{XX XXXXX 2025}
\received[revised]{XX XXXXX XXXX}
\received[accepted]{XX XXXXX XXXX}

\maketitle

\section{Introduction}

A controlled and adjustable release of an active pharmaceutical ingredient (API) from a tablet or implant is essential to ensure proper treatment and patient safety \cite{ref0, REF01}. In the case of tablets, the amount of API in the patient's body as a function of time can be adjusted by tailoring the dosage of the drug; however, this approach is generally not considered a controlled-release mechanism. Significant advancements in drug release control have been made possible through research in pharmaceutical sciences and materials science, which have explored the strategies to create customized tablets and implants that yield sustained, extended, dual, and immediate release, among others \cite{rathna, modeling, controlled}. Treatment of certain diseases, such as cancer, requires a carefully controlled release of the drug \cite{cancer} to maintain a specified amount of the API within the patient's body, whereas treatment of diseases like asthma benefits from drugs that provide pulsatile release profiles \cite{asthma}. Additive manufacturing techniques, such as three-dimensional printing and injection molding, considerably support pharmaceutical perspectives by enabling the production of multi-material drugs and allowing precise structural customization and rapid prototyping \cite{mahran, personalizing, semi, ozliseli}. 

A major difficulty in drug design arises from the wide variety of mechanisms involved in drug release. The drug may release the API through diffusion \cite{diffusion}, erosion \cite{erosion}, dissolution \cite{dissolution}, or a combination of these mechanisms \cite{disso-diff, diff-ero}. The drug can also be subject to phenomena that affect the release rate, such as swelling \cite{cellular}. The various release mechanisms are described in the pharmacy through so-called kinetic models, which can be interpreted as 1-3 parameter functions that can be fitted to experimental data to approximate the drug release profile \cite{kinetics, drug-delivery}. However, kinetic models cannot be applied in the optimal design of drugs since they do not consider important factors such as the shape and material composition of the drug. They are mainly used as tools to identify the release mechanism of a drug, for example, to detect non-Fickian diffusion \cite{non-fiction}. A more mathematical \textit{ab initio} approach for simulating the drug release profile of a given drug composition requires limiting the scope of the model to a specific release mechanism. 

Due to the complex release mechanisms, existing computational drug design approaches typically make several assumptions about the structure of the drug. One way to design the shape of a dissolving or eroding drug is to limit the release process to one dimension only \cite{one-dimentional, multilayers}. This means that the drug is not allowed to dissolve from its sides, in which case the drug consists of a dissolving material coated on its sides with an insoluble and impermeable substance \cite{one-dimentional}. One-dimensional designs are inherently restrictive, as they limit the possible shapes that a drug can take. A different approach was adopted in \cite{parametriccekrepublic}, where the composition of a three-dimensional drug was designed using an evolutionary optimization algorithm. The release model in \cite{parametriccekrepublic} assumes the drug releases the API via dissolution, which takes place on the boundary of the drug. A cylindrical drug is first parameterized into segments with a given dissolution rate. The dissolution process of the segments is simulated using a queue-based, inherently discrete algorithm, which is not based on the discretization of a continuous dissolution model, such as a partial differential equation. As the dissolution proceeds and a segment dissolves, the dissolution continues to the adjacent unsolved segment. The objective is to determine the optimal dissolution rate for each segment among several available dissolution rates, each representing a possible material. The design variables in the method are discrete, which poses significant challenges in the optimization stage due to the problem being a combinatorial optimization problem \cite{discrete}.

Parametric and discrete structural optimization methods typically offer limited design flexibility due to their predefined design spaces. In contrast, topology optimization \cite{bendsoe1995optimization}, being non-parametric, provides greater freedom by enabling the optimization of the entire structure with respect to a performance criterion \cite{allaire2021shape}. As a result, topology optimization is widely regarded as the industry standard in structural design. For example, topology optimization is used to optimize the mass of mechanical devices while maintaining their stiffness properties \cite{stiffness, ECMS}, minimize the compliance of dental prostheses \cite{altunay2025denture}, and improve the efficiency of heat exchangers \cite{mechanical1, DTU-thesis}.

One of the most common topology approaches is the level-set-based method, which considers the boundary of the object as the zero level-set of a signed distance function \cite{allairelevelset}. During the optimization,  the level-set equation is solved to generate new structures \cite{allaire2021shape}. A crucial part of the level-set method is the formulation of the velocity function in the level-set equation, which depends on the objective function of the topology optimization problem. The level-set method can be challenging to implement numerically, especially when the level-set equation and the objective function are solved on different computational grids. However, the level-set method offers high-quality topologies due to its ability to represent sharp interfaces. The second standard method is based on continuous relaxation, meaning that the material distribution function is allowed to take continuous values rather than discrete ones \cite{sigmund2013topology}. A method called Solid Isotropic Material with Penalization (SIMP) belongs to the class of these approaches \cite{originalSIMP}. The continuous relaxation must be used in conjunction with suitable filtering, which aims to project the continuous material distribution to binary designs \cite{regularization2}. The filtering is somewhat heuristic, but the continuous relaxation is straightforward to implement for almost any topology optimization problem and typically yields good results in practice. 

Our primary objective in this paper is to extend the existing methodology and apply topology optimization using continuous relaxation to design drug compositions. Topology optimization is a crucial component of our method, as we aim to avoid relying on parametric methods or those with discrete design variables. In \cite{shapeadjoint}, the topology optimization and adjoint sensitivity analysis were used to design single-material geometries for desired dissolution profiles. The paper proposes the application of the Eikonal equation in modeling dissolving objects. The framework allows for the gradient-based complete topology optimization of 3D-printable structures \cite{shapeadjoint, Jari2023}. However, drug design applications almost always require multi-material structures to achieve the desired release profiles. Unlike in \cite{shapeadjoint}, we must also impose shape constraints on the drug, as they are subject to regulation \cite{drugsize}. Complex drug shapes are undesirable in pharmaceutical applications; for example, tablets must be edible.

A considerable factor in the structural design is the uncertainty of material parameters or boundary conditions. A topology optimization that addresses the uncertainties is known as robust topology optimization (RTO) \cite{sigmund2009manufacturing}. In the context of drug design, uncertainties can be attributed to dissolution rates being random variables or manufacturing defects, for instance. One of our objectives is to apply robust topology optimization methods in the design of drug composition. Robust topology optimization is generally more complex than classical topology optimization, as it requires considering the entire distribution of the random variables in the objective function when determining the final design \cite{robustreview}. In this work, we apply the continuous relaxation-based topology optimization and the stochastic reduced order model (SROM) \cite{SROMgrigoriu} for the inverse design of multi-material controlled-release drugs under uncertainty. The SROM is essentially an uncertainty quantification method that uses a set of carefully selected weighted samples for uncertainty propagation in the forward model. Unlike Monte Carlo methods, SROM requires far fewer samples to function effectively and is not affected by sampling noise. Our overall design method is inspired by the geometric dissolution model of \cite{shapeadjoint}, which we extend to multi-material drugs with random dissolution rates.

Our contributions can be summarized as follows.
\begin{itemize}
    \item We implement a continuous relaxation-based topology optimization method to find the optimal material composition of a drug when the materials have different dissolution rates and concentrations. The method finds the composition that results in a release profile that is close to the target release profile in terms of squared release profile difference. The method is non-parametric, applicable to drugs with an arbitrary shape, and assumes that the release of the drug occurs from its dissolving surface. 
    \item We also account for the uncertainty in dissolution rates during the optimization of drug composition. We apply the SROM to minimize the expected value of the squared release profile differences, which is a stochastic function. To the best of our knowledge, robust topology optimization has not been utilized before in the design of drugs.
\end{itemize}

The rest of the paper is structured as follows. In \Cref{methodsection}, we introduce the optimization method and its uncertainty-aware generalization for drug design. In \Cref{numerialsection}, the proposed methods are applied in selected numerical examples, in which we design the compositions of a capsule-shaped drug having different target release profiles. We conclude our contributions and findings of our work in \Cref{conclusionsection}.  

\section{Method}
\label{methodsection}

To establish a computationally effective model for simulating the drug release, we assume that the dissolution process of the drug follows the Noyes-Whitney law \cite{Noyes-Whitney, Noyes-Whitney-model}. The dissolving drug is assumed to release the API only from its surface. The premise of the Noyes-Whitney drug is that the release rate $ \frac{dC}{dt} $ of the API of the drug is proportional to the surface area of the drug exposed to the solvent. The release rate can then be formulated as 
\begin{equation*}
    \frac{dC}{dt} = vA(C_0-C),
\end{equation*}
where $C$ is the concentration of the API in the solution, $C_0$ is the saturation solubility of the API, $A$ is the surface area of the drug, and $v$ is proportional to the dissolution or reaction rate of the drug embedded in the surrounding solvent. If the container volume is assumed to be very large, the release rate is proportional to the product of the surface area of the drug and its dissolution rate. The mathematical treatment of the release then reduces to modeling the time-dependent boundary of the drug, which is determined by its material composition. This means that we can consider the dissolution rate as a location-dependent function ${v(\bm{x})}$, which does not depend on time-dependent environmental factors such as the solvent concentration or pH. The same modeling principle is used in \cite{shapeadjoint}, which we extend to the computational drug design. Limiting the scope to surface-releasing drugs is essential to enable the use of efficient numerical tools for computational inverse design. 

Our objective in the computational design of the personalized drug is to minimize the misfit functional $J(\rho$), which measures the overall discrepancy between the desired remaining mass of the drug, $R(t)$, and the remaining mass of the designed drug, $R_d(t;\rho)$. The drug material composition depends on the design function $\rho$. The objective is to find such a design function $\rho: \Omega \rightarrow [0,1]$ which minimizes the misfit functional
\begin{equation}
\label{misfit}
     \min_{\rho} J(\rho) = \min_{\rho} \int_0^{\tau_{\textrm{end}}} \left( \frac{R(t)}{R(0)} - \frac{R_d(t;\rho)}{R_d(0;\rho)} \right)^2 dt,
\end{equation}
where $\tau_{\textrm{end}}$ denotes the end time of the desired drug release. The misfit functional minimizes the squared differences of the normalized release profiles, assuming that $R(0) > 0$ and $R_d(0;\rho) > 0$ hold always. We are not interested in the absolute magnitudes of the drugs' release profiles but rather in the normalized release profiles because the absolute magnitude can be easily adjusted by dosage or scaling the concentrations of the API in the drug. Normalization can also improve flexibility in the design space by increasing the degrees of freedom in the designs.
The design function determines the distribution of materials inside the drug. When $\bm{x}=0$, the material at $\bm{x}$ should consist of material number one. Similarly, when $\rho(\bm{x}) = 1$, material number two is assigned to $\bm{x}$. We employ continuous relaxation in the drug design, meaning that the design function can take intermediate values between 0 and 1 \cite{sigmund2013topology}. The value of $\rho$ represents simply the fractional composition of the materials. For instance, $\rho(\bm{x}) = 0.7$ indicates that the material at position $\bm{x}$ is composed of 70\% of the second material and 30\% of the first material. Thanks to continuous relaxation, the sensitivities of the loss function with respect to the design variables can be established. 

Based on these assumptions, we can use the Eikonal equation as the governing equation for the drug release process \cite{shapeadjoint}. The Eikonal equation is used predominantly in geometrical optics to model the wavefront propagation \cite{eikonalinfo2}. It is also applied in the estimation of seismic velocity and hypocenter location in computational seismology \cite{eikonalinfo, eikonalinfo3}. The solution of the Eikonal equation $T(\bm{x})$ equals the minimal time when a propagating wavefront reaches $\bm{T}$ \cite{sethian1999level}. In our case, we interpret the surface of the drug as a virtual wavefront.  Given the normal velocity function $v(\bm{x})$ and the initial drug shape, the boundary of the dissolving drug at all times can be found by solving the Eikonal equation only once. The Eikonal equation is
\begin{equation}
\label{eiko}
\begin{alignedat}{2}
    &\Vert \nabla T(\bm{x}) \Vert = \frac{1}{v(\bm{x})}, \quad &&\bm{x}  \in \Omega,\\
    &T(\bm{x}) = 0, \quad &&\bm{x}\in \partial \Omega
\end{alignedat}
\end{equation}
where $\Omega \subset \mathbb{R}^3$ is the domain of the drug at, $\partial \Omega $ is the surface of the drug before the dissolution starts, and $\Vert \cdot \Vert$ stands for the Euclidean norm.

The relationship between the solution of the Eikonal equation and the design release profile $R_d(t;\rho)$ is straightforward, as it is derived directly from the Eikonal equation itself. The mass of the drug at time $\tau$ equals the mass of the drug parts for which $ T(\bm{x}) - \tau  \geq 0$. The mass is simply the integral of the concentration and the Heaviside function over $\Omega$:
\begin{equation}
\label{integra}
R_d(t;\rho)   = \int_{T(\bm{x}) - t \geq 0} c(\rho) d\bm{x} = \int_{\Omega} c(\rho) H\left(T(\bm{x})-t\right)d\bm{x},
\end{equation}
where $c(\rho)$ is the concentration function, and $H$ is the Heaviside step function. For simplicity, we assume both $c$ and $v$ depend linearly on the design function $\rho$:
\begin{alignat}{1}
\label{equs}
    &v(\rho) = v_1 + (v_2-v_1)\rho(\bm{x})\\
    &c(\rho) = c_1 + (c_2-c_1)\rho(\bm{x}),
\end{alignat}
where $v_1,v_2$ and $c_1,_c2$ are the velocities and concentrations of materials one and two, respectively. In our case, the velocities mean the dissolution rates of the materials. 
Although the continuation relaxation is used, the design function $\rho$ should be as binary as possible, meaning that it takes only values close to zero or one. In topology optimization, the binary designs are achieved through regularization and filtering techniques \cite{regularization, regularization2, densityfilter}. In the course of optimization, we apply a density filter and a smooth Heaviside filter on the design function $\rho$ \cite{smoothHeavisidefunction}. The application of the averaging filter can be represented as a convolution
\begin{equation}
\label{convolution}
\hat{\rho}(\bm{x}) = (\phi * \rho)(\bm{x}) = \int_{\Omega} \phi(\bm{x}-\bm{y})\rho(\bm{y}) d\bm{y},
\end{equation}
where $\phi$ stands for the averaging kernel. The kernel can be, for example, a linearly decaying averaging function \cite{sigmund2009manufacturing}. 
 The density filter regularizes the design function, promoting designs that avoid small structural features \cite{densityfilter}. 
The density filter is typically used in conjunction with the smooth Heaviside filter, which aims to project the averaged density back to close-to-binary design variables after the density filter has been applied. The smooth Heaviside filter is used point-wise to the density-filtered design $\hat{\rho}(\bm{x})$ \cite{smoothHeavisidefunction} to yield the final design $\overline{\rho}(\bm{x})$, which is used in the objective functional instead of unfiltered $\rho$. There are several different smooth Heaviside filters \cite{regularization2}, but one of the most common ones is based on the hyperbolic tangent functions, which is of the form
\begin{equation}
\label{heaviside}
\overline{\rho}(\bm{x}) = \frac{\tanh(\beta/2) - \tanh\left(\beta(\hat{\rho}(\bm{x}) - 1/2 \right) ) }{2\tanh(\beta/2)}.
\end{equation}
In \Cref{heaviside}, parameter $\beta$ is the filtering parameter. The greater the $\beta$, the steeper the filter is, meaning that the filter approaches the exact Heaviside filter when $\ \beta\rightarrow\infty$. 


\subsection{Discretization}
The integral in the misfit functional in \Cref{misfit}  is approximated as a sum in the discretized objective function, which is
\begin{equation}
\label{discremisfit}
 \min_{\bm{\rho}} J(\bm{\rho})  =  \min_{\bm{\rho}}  \sum_{i=1}^{M} \left( \frac{R(t_i)}{R(0)} - \frac{R_d(t_i;\overline{\bm{\rho}})}{R_d(0;\overline{\bm{\rho}})} \right)^2 \Delta t_i,
\end{equation}
where $M$ is the number of time points in the discretization of the release profile time series and $\Delta t_i$ is the interval between the adjacent time points. In our numerical experiments, we assume $\Delta t_i = \textrm{constant}$. We recall that the volume of the drug is computed using the density filtered and Heaviside projected $\overline{\bm{\rho}}$, and that is why the mass evaluation functions $R_d$ use the filtered design vectors. The Eikonal equation of the dissolution model in \Cref{eiko} is discretized in a Cartesian grid and solved through the first-order fast marching method (FMM) \cite{sethian1999level}. The first-order FMM iteratively solves the equations 
\begin{equation}
\label{fmm}
\begin{split}
&\max\left(\frac{T_{i,j,k} - \min(T_{i+1,j,k}, T_{i-1,j,k})}{h_x},0 \right)^2 +  \\&\max\left(\frac{T_{i,j,k} - \min(T_{i,j+1,k}, T_{i,j-1,k})}{h_y},0 \right)^2 + \\ &\max\left(\frac{T_{i,j,k} - \min(T_{i,j,k+1}, T_{i,j,k-1})}{h_z},0 \right)^2 - \frac{1}{v_{i,j,k}^2} = 0
\end{split}
\end{equation}
for $1\leq i \leq N_i$, $1\leq j \leq N_j$, and $1\leq k \leq N_k$  for the discretized time-to-reach function values $T_{i,j,k}$. In \Cref{fmm}, the Cartesian grid intervals in X, Y, and Z directions are $h_x$, $h_y$, and $h_z$, respectively. In practice, the equations can be solved with the help of a priority queue. The FMM initiates the iterations from grid points that are at or near the boundary where $T_{i,j,k} = 0$, and labels those grid points as accepted. The points adjacent to the accepted points are labeled as considered and pushed to the priority queue along with their tentative $T_{i,j,k}$ values. The algorithm proceeds then by popping the grid point with the lowest value $T$ from the queue, setting it as a fixed grid point, and recomputing the $T_{i,j,k}$ values of the adjacent points of the new fixed point. The adjacent points are then pushed to the queue if they are not there already; otherwise, their $T$ values are updated.  The iterative priority queue-based procedure is possible thanks to the upwind finite-difference scheme, which makes it $T_{i,j,k}$ depend on its adjacent neighbor values through a causal connection \cite{sethian1999level}. There are also higher-order FMM implementations, such as second-order schemes \cite{secondorderFMM} and the multi-stencil FMM \cite{multistencilsFMM}. The second-order scheme offers higher-order convergence with respect to the Cartesian grid size, while the multi-stencil FMM reduces the anisotropy of the discretized Eikonal equation.

The priority queue makes the FMM a relatively efficient method for solving the Eikonal equation on Cartesian grids, with a computational complexity of $\mathcal{O}(N \log N)$, where $N$ is the number of degrees of freedom in the discretization \cite{sethian1999level}. In contrast to other classical numerical methods to solve the Eikonal equation, the FMM is straightforward to implement, and its computational cost does not depend on the dissolution rates $v(\bm{x})$ \cite{FMMvsFSM}. The algebraic equations of the first-order FMM are also differentiable, which enables the computation of partial derivatives of the discretized time-to-reach function $\bm{T}$ with respect to the discretized dissolution rate $\bm{v}$ on the Cartesian grid. Due to these features, the first-order FMM is our selected method for solving the Eikonal equation in the optimal design problem, similar to \cite{shapeadjoint}. For more information regarding the FMM, we refer to \cite{sethian1999level}. 

The FMM is used in a Cartesian grid $\mathcal{G} \subset \mathbb{R}^3$, so the domain of the drug $\Omega \subset \mathcal{G}$. A small modification must be made in \Cref{integra} to permit the usage of the Cartesian grid $\mathcal{G}$ as the computational grid. We employ the sign of the signed distance function $\phi(x,y,z)$, which tells the signed distance to the boundary $\partial\Omega$ of the drug. That is, 
\begin{equation*}
\textrm{sign}(\phi(x,y,z))=    \begin{cases}
			-1, & (x,y,z) \in \Omega \\
            0, & (x,y,z) \in \partial \Omega \\
            1, & (x,y,z) \notin \Omega.
		 \end{cases}
\end{equation*}
As the discrete counterpart of the distance filtering of \Cref{convolution}, we apply a normalized weighted sum of the form
\begin{equation}
\label{density}
\hat{\rho}_{i,j,k} = \frac{\sum_{a,b,c} \max\left(0,d_{\text{max}}-\Vert\bm{x}_{i,j,k} - \bm{x}_{a,b,c}\Vert\right)\rho_{a,b,c} }{\sum_{a,b,c}^N\max\left(0,d_{\text{max}}-\Vert\bm{x}_{i,j,k} - \bm{x}_{a,b,c}\Vert\right)},
\end{equation}
for each design variable $\rho_{i,j,k}$ in the grid, where $d_{\text{max}}$ denotes the characteristic length \cite{densityfilter}. In practice, density filtering is implemented as a matrix $\bm{D}$, where the entries of the matrix are computed only once before being used in the filtering process.  The smooth Heaviside projection complements the filtering of the discrete design variables by directly applying the Heaviside approximation from  \Cref{heaviside}  to the distance-filtered design variables.

Instead of attempting to approximate the integral of the discrete time-to-reach function $\bm{T}$ in the implicitly defined domain $\Omega$, we approximate the  integral \Cref{integra} with a sum
\begin{equation}
\label{approxi}
\begin{split}
    & R_d(t;\bm{\rho})  = \\ & h_x h_y h_z \sum_{i=1}^{N_i-1} \sum_{j=1}^{N_j-1} \sum_{k=1}^{N_k-1} \textsf{mean}(i,j,k;\bm{c}) \mathsf{Vol}(i,j,k; \bm{s} \odot \bm{T} + t ), 
    \end{split}
\end{equation}
where $ \odot$ stands for the Hadamard product. $N_i$, $N_j$, and $N_k$ are the number of grid points in the X, Y, and Z axes, respectively.  The total approximation of the drug mass at time $t$ is computed as a sum over the grid cells since the Eikonal equation is solved on the grid nodes. The approximation is given by \begin{equation}
\label{avg}\textsf{mean}(i,j,k;\bm{c}) =  \frac{1}{8} \left(\sum_{d_x, d_y,d_z = \{ 0,1\} } c_{i+d_x,j+d_y,k+d_z}\right).
\end{equation}
In \Cref{approxi}, $\mathsf{Vol}(i,j,k; \bm{F}) $ is a function relying upon the marching cubes algorithm that computes the volume fraction of the Cartesian grid cell having the vertices of $(i,j,k)$, $(i+1,j,k)$, $i,j+1,k)$ and $(i,j,k+1)$ where the discretized scalar quantity $\bm{F}$ is negative. More specifically, it $\mathsf{Vol}(i,j,k; \bm{F}) $ implements a marching cubes-based method to compute the volume fraction \cite{takahashi2022fast}. Unlike most other implicit domain integration methods, this method is virtually exact when the discretized scalar quantity $\bm{F}$ is interpolated linearly within the cell. The method is also computationally very efficient.  Finally, $\bm{s}$ denotes the sign of the signed distance function evaluated in all Cartesian grid points: $s_{i,j,k} =  \textrm{sign}(\phi(x_{i,j,k},y_{i,j,k},z_{i,j,k}))$. 

\subsection{Gradient computation}

The motivation of the continuous relaxation method is to enable the evaluation of the gradients $\frac{\partial J(\bm{\rho})}{\partial \bm{\rho}}$ from \Cref{discremisfit} for optimization algorithms like the quasi-Newton methods. The gradient evaluation of the objective function is the most challenging for the mass evaluation terms \Cref{approxi}. $R_d$ is a function of both the discretized concentration ${\bm{c}}$ and the FMM solution $\bm{T}$. The concentration depends directly on $\bm{\rho}$. $\bm{T}$ depends on the discretized velocity $\bm{v}$, which of course also depend of $\bm{\rho}$. Both of the dependencies are given in \Cref{equs}. To simplify the notation, we can write the objective function as a function of the filtered design variable $\overline{\bm{\rho}}$ as follows:
\begin{equation*}
     J(\overline{\bm{\rho}}) =  J(\bm{T}(\bm{v}(\overline{\bm{\rho}})), \bm{c}(\overline{\bm{\rho}})).
\end{equation*}
The mass evaluation function can also be written similarly:
\begin{equation*}
     R_d(t;\overline{\bm{\rho}}) =  R_d(t; \bm{T}(\bm{v}(\overline{\bm{\rho}})), \bm{c}(\overline{\bm{\rho}})).
\end{equation*}
The gradient of the mass evaluation function $R_d$ with respect to $\overline{\bm{\rho}}$ requires hence finding 
\begin{equation*}
\frac{\partial R_d(t; \bm{T}(\bm{v}(\overline{\bm{\rho}})), \bm{c}(\overline{\bm{\rho}})) }{\partial\bm{T} } \frac{\partial \bm{T} }{\partial{\bm{v}} }
\end{equation*}
and 
\begin{equation*}
\frac{\partial R_d(t; \bm{T}(\bm{v}(\overline{\bm{\rho}})), \bm{c}(\overline{\bm{\rho}}))  }{\partial\bm{c} } 
.
\end{equation*}
The gradient $\frac{\partial R_d(t; \bm{T}(\bm{v}(\overline{\bm{\rho}})), \bm{c}(\overline{\bm{\rho}}))  }{\partial\bm{c} }  $ can be calculated analytically from \Cref{approxi} by noticing that the concentration $c_{i,j,k}$ contributes to the total sum  at most eight times in \Cref{approxi} and the term $ \mathsf{Vol}(i,j,k; \bm{s} \odot \bm{T} + t  )$ does not depend on $\bm{c}$. Hence, the gradient can be computed simultaneously during the evaluation of \Cref{approxi} over all grid cells.

The gradient $ \frac{\partial R_d(t; \bm{T}(\bm{v}(\overline{\bm{\rho}})), \bm{c}(\overline{\bm{\rho}}))  }{\partial\bm{T} } $ is a bit trickier to compute. The marching cubes integration method defines 23 different cases in $\mathsf{Vol}(i,j,k; \bm{F}) $ to evaluate the volume where a given signed distance function is negative \cite{takahashi2022fast}. The 23 cases are the result of a manual construction of all possible ways that the marching cubes algorithm can split a cube into disjoint parts having different signs of the discretized signed distance function $\bm{F}$ \cite{takahashi2022fast}. The analytical gradients for the volume evaluation algorithm with respect to the signed distance function would require major bookkeeping and reorganization of the code. However, the gradients can be approximated rather efficiently using the finite difference method. $T_{i,j,k}$, the value of $T$ at node $(i,j,k)$ contributes to the signed distance function volume computation at most eight adjacent cells, so the finite-difference approximation for  $  \frac{\partial R_d(t; \bm{T}(\bm{v}(\overline{\bm{\rho}})), \bm{c}(\overline{\bm{\rho}}))  }{\partial\bm{T} }  $  is computed in the same time as the computation of $\frac{\partial R_d(t; \bm{T}(\bm{v}(\overline{\bm{\rho}})), \bm{c}(\overline{\bm{\rho}}))  }{\partial\bm{c} }$  during the evaluation of \Cref{approxi}. The approach requires a small amount of additional bookkeeping, but the approximation remains accurate and reliable. The volume of the negative signed distance function parts inside a cell equals the volume of the whole cell when the signed distance function is negative in all the vertices of the cell regardless of their magnitudes, so the derivative contribution of the $T_{i,j,k}$ in these all-negative cells can be directly set to zero.

The computationally hardest part is to compute $ \frac{\partial \bm{T} }{\partial\bm{v }} $. We employ the discrete adjoint method for this purpose  \cite{adjoexample}, and the approach is adapted from the geometrical shape design from \cite{shapeadjoint}. The rationale of the method is to differentiate the system of the nonlinear equations of the FMM \Cref{fmm} with respect to $\bm{T}$ and $\bm{v}$. Let $\bm{E}(\bm{T},\bm{v})$ stand for the FMM equations, so $E_{i,j,k}$ denotes the equation for the $(i,j,k)^{\text{th}}$ index. Differentiating $\bm{E}$ yields a new equation
\begin{equation}
\label{deriva}
   \frac{\partial \bm{E}}{\partial \bm{T}} \frac{\partial \bm{T}}{\partial {\bm{v}}} + \frac{\partial \bm{E}}{\partial \bm{v}} 
   = \bm{0}.  
\end{equation}
The interesting part in \Cref{deriva} is the term $\frac{\partial \bm{T}}{\partial {\bm{v}}}$.
Solving \Cref{deriva} with respect to $\frac{\partial \bm{T}}{\partial \overline{\bm{\rho}}}$ would give us
\begin{equation*}
    \frac{\partial \bm{T}}{\partial {\bm{v}}} = - \left(\frac{\partial \bm{E}}{\partial \bm{T}}\right)^{-1} \frac{\partial \bm{E}}{\partial \bm{v}}, 
\end{equation*}
but doing so would require inverting the Jacobian matrix $\frac{\partial \bm{E}}{\partial \bm{T}}$, which is virtually impossible due to the memory requirements. Instead, $\frac{\partial \bm{T}}{\partial {\bm{v}}} $ can be directly inserted into  $\frac{\partial J(\bm{T}(\bm{v}(\overline{\bm{\rho}})), \bm{c}(\overline{\bm{\rho}}))}{\partial \bm{T}} \frac{\partial \bm{T}}{\partial {\bm{v}}} $, 
which yields
\begin{equation}
\label{adjoidea}
\begin{split}
& \frac{\partial J(\bm{T}(\bm{v}(\overline{\bm{\rho}})), \bm{c}(\overline{\bm{\rho}}))}{\partial \bm{T}}\frac{\partial \bm{T} }{\partial{\bm{v}} } = \\&-\frac{\partial J(\bm{T}(\bm{v}(\overline{\bm{\rho}})), \bm{c}(\overline{\bm{\rho}}))}{\partial \bm{T}}\left(\frac{\partial \bm{E}}{\partial \bm{T}}\right)^{-1} \frac{\partial \bm{E}}{\partial \bm{v}}. %
\end{split}
\end{equation}
The matrix inversion can be avoided by solving the adjoint equation
\begin{equation}
    \label{adjoint}
     \left(\frac{\partial \bm{E}}{\partial \bm{T}}\right)^{\top}\bm{q} = \frac{\partial J(\bm{T}(\bm{v}(\overline{\bm{\rho}})), \bm{c}(\overline{\bm{\rho}})) }{\partial\bm{T} }.
\end{equation}
In \Cref{adjoint}, $\frac{\partial J(\bm{T}(\bm{v}(\overline{\bm{\rho}})), \bm{c}(\overline{\bm{\rho}}))}{\partial \bm{T}} $  is given by
\begin{equation*}
\begin{split}
     &\frac{\partial J(\bm{T}(\bm{v}(\overline{\bm{\rho}})), \bm{c}(\overline{\bm{\rho}}))}{\partial \bm{T}} = \\ & \sum_{i=1}^M -2\Delta t_i \left( \frac{R(t_i)}{R(0)} - \frac{R_d(t_i;\overline{\bm{\rho}})}{R_d(0;\overline{\bm{\rho}})} \right) \frac{    R_d(0;\overline{\bm{\rho}}) \bm{h}_i \odot \bm{s} -   R_d(t_i;\overline{\bm{\rho}})  \bm{h}_0 \odot \bm{s} }{R_d(0;\overline{\bm{\rho}})^2} ,
     \end{split}
\end{equation*}
where $ \bm{h}_i =  \frac{\partial  R_d(t_i; \bm{T}(\bm{v}(\overline{\bm{\rho}})), \bm{c}(\overline{\bm{\rho}})) }{\partial\bm{T} }$.
The solution can then be inserted back into \Cref{adjoidea}, giving
\begin{equation*}
\frac{\partial J(\bm{T}(\bm{v}(\overline{\bm{\rho}})), \bm{c}(\overline{\bm{\rho}})) }{\partial\bm{T} } \frac{\partial \bm{T} }{\partial { \bm{v} } } = -\bm{q}^{\top} \frac{\partial \bm{E}}{\partial \bm{v}}.
\end{equation*}

Finally, the gradient of \Cref{discremisfit} with respect to the unfiltered design variables $\bm{\rho}$ can now be written with the help of the chain rule and the adjoint equation solution as
\begin{equation}
\label{finalgrad}
\begin{split}
    \frac{\partial J(\bm{\rho})}{\partial \bm{\rho}}& = (v_2 - v_1) \bm{D}^{\top}\textrm{vec} \left(\frac{\partial \overline{\bm{\rho}}}{\partial \hat{\bm{\rho}}}  \left( -\bm{q}^{\top} \frac{\partial \bm{E}}{\partial \bm{v}} \right) \right)\\
    %
    %
     &+ (c_2- c_1)\bm{D}^{\top}\textrm{vec}\left(\frac{\partial \overline{\bm{\rho}}}{\partial \hat{\bm{\rho}}} \frac{\partial J(\bm{T}(\bm{v}(\overline{\bm{\rho}})), \bm{c}(\overline{\bm{\rho}}))}{\partial \bm{c}}   \right). \end{split}
\end{equation}
In \Cref{finalgrad}, $\frac{\partial J(\bm{T}(\bm{v}(\overline{\bm{\rho}})), \bm{c}(\overline{\bm{\rho}}))}{\partial \bm{c}} $ is given by 
\begin{equation*}
\begin{split}
     &\frac{\partial J(\bm{T}(\bm{v}(\overline{\bm{\rho}})), \bm{c}(\overline{\bm{\rho}}))}{\partial \bm{c}} = \\
     &\sum_{i=1}^M -2\Delta t_i \left( \frac{R(t_i)}{R(0)} - \frac{R_d(t_i;\overline{\bm{\rho}})}{R_d(0;\overline{\bm{\rho}})} \right) \frac{   R_d(0;\overline{\bm{\rho}}) \bm{k}_i -   R_d(t_i;\overline{\bm{\rho}})   \bm{k}_0}{R_d(0;\overline{\bm{\rho}})^2} ,
     \end{split}
\end{equation*}
with $ \bm{k}_i =  \frac{\partial R_d(t_i; \bm{T}(\bm{v}(\overline{\bm{\rho}})), \bm{c}(\overline{\bm{\rho}}))  }{\partial\bm{c} }$.  
This completes the gradient evaluation for $\frac{\partial J(\bm{\rho})}{\partial \bm{\rho}}$.



\subsection{Robust topology optimization}

In classical topology optimization, the material parameters and boundary conditions are assumed to be known and fixed. In our case, the material parameters include the material-specific dissolution rates and the concentrations of the drug. When some of the drug parameters are random variables rather than fixed ones, considering a misfit function like in \Cref{misfit} as the objective function is inappropriate, as these random parameters produce a distribution of release profiles rather than a single deterministic release curve. More generally, even the final design can be interpreted as random due to manufacturing errors and imperfections. Addressing the uncertainties in the optimal drug design requires an alternative approach. 

The robust topology optimization (RTO) builds upon the methodology from robust design optimization (RDO) and reliability-based (robust) design optimization (RBDO, RBO, or RBRDO if used in conjunction with the robust design optimization) to account for the uncertainty in the design and to maintain a predictable performance in the presence of the uncertain parameters \cite{robustreview, robustreview2, rbrdo}. In the RBDO, the premise is to set a reliability measure as a constraint to minimize a certain objective function \cite{rbdo}. The reliability measure is typically a lower limit for the probability that a desired condition occurs \cite{rbdo,robustreview}. Adapted to the case of drug design, the objective function would still be of the form \Cref{misfit}, but the values of the objective function should not exceed a certain threshold with a given probability. Hence, the RBDO considers probabilities as design constraints. The principle is concrete from a reliability engineering perspective, but it is often challenging to implement, as the probabilities must be computed either through Monte Carlo methods \cite{robustreview} or analytic approaches, such as the first- and second-order reliability methods \cite{formsorm}. It can also be argued that setting a threshold for the probability is not always obvious.

The RDO does not incorporate probability constraints into the optimization process; instead, it covers three different approaches to account for uncertainty. First, the method can optimize for the worst-case scenario. The objective is to find such design parameters for which the supremum of the objective function with respect to the random variables is the smallest \cite{robustreview}. The minimization problem is, therefore, a min-max problem, which requires defining a feasible set for the random parameters where the supremum of the objective function for a given design is computed \cite{robustreview}. The worst-case approach has been used to address manufacturing defects in topology optimization \cite{tolerantdesign}. The approach lacks a probabilistic interpretation, as it does not account for the probability distribution or variability of the objective function induced by the random parameters. Optimizing for the worst-case scenario is certainly a plausible approach to establishing robust designs, but it may be too prohibitive if most manufactured drugs still deviate significantly from the target release profile. For instance, minimizing the worst-case misfit of the designed drug release profile and the target release profile may be good from a safety perspective. 

Robust design optimization can also involve maximizing a specified design probability \cite{robustreview}. While somewhat related to RBDO, this approach involves a probability measure in the optimization; however, unlike RBDO, probabilities are indeed maximized rather than being incorporated as constraints. The probability maximization approach is somewhat more attractive than the worst-case approach, as it considers the full distribution of random variables; however, it is rarely used in practice due to various reasons. The computation of probabilities requires resorting to approximations, such as the Monte Carlo method, since the probabilities are intractable, and the selection of the probability measure itself may be unclear \cite{robustreview}. In optimal drug design, this approach can be interpreted as maximizing the probability that the misfit between the designed and target release profiles remains below a specified threshold. 

The third RDO technique is based on the expectancy measures \cite{robustreview}. The principle is to minimize an objective function, which is typically either an expected value of an objective function or a weighted sum of the expected value and the standard deviation of an objective function \cite{robustreview}. Minimizing the sum of the expected value and variance of the objective function addresses some of the limitations of other RDO approaches. Unlike in probability maximization, any heuristic probability thresholds are not needed. Compared to the worst-case approach, the expectancy measure approach still considers the full distribution of random variables. The expectancy measure approach defines a multi-objective optimization problem where the weights for expected value and standard deviation reflect their relative importance. The major challenge remains the numerical evaluation of expected values and variances, which must be performed through Monte Carlo or related methods that propagate uncertainty \cite{robustreview}. The expectancy measure RDO has been successfully employed in various topology optimization problems \cite{robustexample, torres2021robust}.

For these reasons, we employ the expectancy measure approach in the uncertainty-aware inverse design of drugs. We minimize the sum of the expected sum of the misfit $\mathbb{E}\left[J(\bm{\rho}; \bm{\xi})\right]$, the squares of release profile differences between design and target release profiles, and the weighted standard deviation $\sqrt{\mathbb{V}\left[J(\bm{\rho}; \bm{\xi})\right]}$ of the misfit. The robust topology optimization objective function is
\begin{equation}
\label{sromobj}
  \mathbb{E}\left[J(\bm{\rho}; \bm{\xi})\right] + k\sqrt{\mathbb{V}\left[J(\bm{\rho}; \bm{\xi})\right]} \rightarrow \min,
\end{equation}
and it is minimized with respect to $\bm{\rho}$. In \Cref{sromobj}, $J(\bm{\rho}; \bm{\xi})$ is the objective function from \Cref{discremisfit}, and $\bm{\xi}$ denotes the random variables that are implicitly part of $J$. In our case, we consider the dissolution rates $v_1$ and $v_2$ of the materials as random variables having a certain probability distribution, so $\bm{\xi} =(v_1, v_2)$. The expected value and the standard deviation terms of the objective function are intractable, so we need to employ approximations to evaluate the objective function and its gradient. Instead of a Monte Carlo-based approach, we prefer the SROM in the approximation, which we introduce in the following section.

\subsubsection{Stochastic Reduced Order method}
In essence, the SROM offers a low-order approximation of the random input variables $\bm{\xi}$ for a stochastic function \cite{SROMgrigoriu}. The rationale behind the method is to propagate uncertainty in a non-intrusive manner through a stochastic function for uncertainty quantification purposes, similar to Monte Carlo methods. Intrusive methods, such as some implementations of the stochastic Galerkin method, necessitate modifications to the forward model in order to compute the output statistics of the stochastic function \cite{sromcompa}. Non-intrusive methods, such as the SROM and the stochastic collocation method, do not require modifying the forward model to incorporate the effect of the random inputs $\bm{\xi}$ into it. In contrast to the stochastic collocation method, the SROM does not need to design an appropriate sparse quadrature with fixed bounds for the random input variables $\bm{\xi}$ to approximate the expected value and the variance in \Cref{sromobj} \cite{sromcompa}.  

The rationale of the SROM is to find a weighted set of representative samples that approximates the target distribution as accurately as possible. This approach has two benefits. First, the uncertainty propagation in the SROM does not introduce sampling noise, as the samples remain constant throughout the stochastic forward model evaluation. Second, it is often sufficient to use fewer samples in the SROM approach than in the Monte Carlo methods because all the samples in the Monte Carlo have the same weight \cite{SROMgrigoriu, torres2021robust} and they are treated as equally important. Hence, it can be argued that the SROM accounts for the tails of the target distribution more carefully than the Monte Carlo methods, thanks to the weighted and optimized sample vectors.

We apply the SROM variant from \cite{sromgrigoriu2} to construct the low-order weighted discrete approximation  $(\bm{w}, \bm{\Psi})$, $\bm{w} \in \mathbb{R}^l$ and $\bm{\Psi} \in \mathbb{R}^{2\times l}$, for the bivariate stochastic dissolution rate vector $\bm{\xi}$. The SROM solves for a constrained minimization problem given by
\begin{equation}
\label{srommi}
\begin{split}
    (\bm{w}, \bm{\Psi}) &= \arg\min_{\bm{w},\bm{\Psi}}\left( \sum_{i=1}^3  \alpha_i h_i(\bm{w}. \bm{\Psi})  \right) \\
    & \textrm{s.t. } \sum_{k=1}^l w_k = 1, \\
    & w_k > 0, \; k = 1, \dots, l.
    \end{split}
\end{equation}
The objective function in \Cref{srommi} consists of three cost functions and the functions are weighted by $\alpha_i > 0$. The function $h_1$ is the cost function for the differences between the target cumulative distribution functions (CDF) $F_i$ and the smoothed SROM CDF approximations, and it is  
\begin{equation*}
\begin{split}
   &h_1(\bm{w}, \bm{\Psi}) = \\ &\frac{1}{2} \sum_{i=1}^2 \sum_{j=1}^l \left( \frac{1}{2} \sum_{k=1}^l w_k \left( 1+ \mathrm{erf}\left( \frac{ \Psi_{i,j}  - \Psi_{i,k}}{\sqrt{2}\sigma}\right ) \right )  -F_i(\Psi_{i,j}) \right)^2,
   \end{split}
\end{equation*}
where $\sigma > 0$ is a regularization parameter to smooth the CDF function.
The second cost function of \Cref{srommi}  aims to match the non-central moments of the SROM approximation to the exact ones:
\begin{equation*}
   h_2(\bm{w}, \bm{\Psi})  = \frac{1}{2} \sum_{i=1}^2 \sum_{m=1}^{W} \left(  \frac{\tilde{\mu}_i(m) - \mu_i(m)}{\mu_i(m)} \right)^2,
\end{equation*}
where $\mu_i(m) = \mathbb{E}[\xi_i ^m]$ and $\tilde{\mu}_i(m) = \sum_{j=1}^l w_j \Psi_{j,i}^m$. 
The remaining cost function $h_3$ seeks to match the correlation matrices of the SROM approximation to the exact one through
\begin{equation*}
   h_3(\bm{w}, \bm{\Psi})  = \frac{1}{2} \sum_{i,j=1, j>1}^2 \left(  \frac{\tilde{r}(i,j) - r(i,j)}{r(i,j)} \right)^2,
\end{equation*}
where $r(i,j) = \mathbb{E}[\xi_i \xi_j]$ and $\tilde{r}(i,j) = \sum_{k=1}^l w_k \Psi_{i,k}\Psi_{j,k}$. 
The uncertainty can be propagated in the forward model $J(\bm{\rho};\bm{\xi})$ by simply evaluating it with all the constructed sample vectors $\bm{\Psi}$ and then weighting the outputs by the weights $\bm{w}$. That is, 
\begin{equation}
\label{expect}
      \mathbb{E}\left[J(\bm{\rho}; \bm{\xi})\right]  \approx \frac{1}{l} \sum_{i=1}^l w_i J(\bm{\rho}; \bm{\Psi}_{:,i}),
\end{equation}
and
\begin{equation}
\label{varia}
\begin{split}
      &\sqrt{\mathbb{V}\left[J(\bm{\rho}; \bm{\xi})\right]}  \approx \\ &\sqrt{\left( \frac{1}{l} \sum_{i=1}^l w_i J(\bm{\rho}; \bm{\Psi}_{:,i})^2 \right) - \left( \frac{1}{l} \sum_{i=1}^l w_i J(\bm{\rho}; \bm{\Psi}_{:,i}) \right)^2}.
      \end{split}
\end{equation}
The gradients of the stochastic objective function in \Cref{sromobj} can be approximated by the SROM samples in a straightforward manner \cite{torres2021robust}:
\begin{equation*}
\begin{split}
    &\frac{\partial \left( \mathbb{E}\left[J(\bm{\rho}; \bm{\xi})\right] + k\sqrt{\mathbb{V}\left[J(\bm{\rho}; \bm{\xi})\right]} \right) }{\partial \bm{\rho}} \\&= \sum_{i=1}^l w_i \frac{\partial J(\bm{\rho}; \bm{\Psi}_{:,i})}{\partial \bm{\rho}} \left( 1 + \frac{k}{\sqrt{v}} \left( J(\bm{\rho}; \bm{\Psi}_{:,i}) - m\right) \right),
    \end{split}
\end{equation*}
where $m = \mathbb{E}\left[J(\bm{\rho}; \bm{\xi})\right]$ from \Cref{expect} and $\sqrt{v} =   \sqrt{\mathbb{V}\left[J(\bm{\rho}; \bm{\xi})\right]}$ from \Cref{varia}.

\section{Numerical examples}
\label{numerialsection}
The numerical experiments with the proposed method are conducted on a rectilinear grid. To promote symmetric designs, we design only one octant of the final drug and make the drug symmetric with respect to the coordinate axis by using the octant design and mirroring. We use $128\times128\times128$ discretization for the FMM. In all our experiments, we design the composition of a capsule drug, which consists of a cylindrical part and has hemispheres at the bottom and top. The radius of the cylinder in the xy-plane is $r=2.32\,\textrm{mm}$, and the overall height of the capsule along the z-axis is $l = 12.49\,\textrm{mm}$. To discretize the octant of the drug, we employ the grid spacings $h_x = h_y = \frac{2.35}{128}\,\textrm{mm}$ for the x- and y-axes and $h_z = \frac{6.25}{128}\,\textrm{mm}$ for the z-axis. We select this grid discretization since it is a good compromise for computational cost and resolution. We construct the signed distance function for the drug. This signed distance function is used to set the boundary condition for the Eikonal equation by identifying the grid nodes that lie on the surface of the drug and the nearest grid points adjacent to it.  

We use the bounded limited-memory Broyden-Fletcher-Goldfarb-Shanno algorithm (L-BFGS-B) as an optimizer algorithm to minimize the objective function \cite{optimizer}. We use natural bounds of zero and one for all $128^3=2097152$ design variables. We use the continuation scheme in the optimization to favor the existence of close-to-binary material compositions \cite{continuationscheme}.
This means that the objective functions are minimized as the smooth Heaviside filtering \Cref{heaviside} increases in value. After the former optimization with a smaller $\beta$ has stopped due to tolerance conditions, the minimization is restarted with a larger $\beta$ in the Heaviside filtering, using the final design parameters from the former round as the initial point for the next optimization round. The procedure is performed for eight rounds, and we use the values $(1, 5, 10,  20, 35, 50, 150, 300)$ for the $\beta$. The discretization and the optimization settings are identical for deterministic and robust designs. The only difference is their objective functions. We recall that for the deterministic designs, we minimize the mean squared differences of the target release profile and the normalized design release profile \Cref{misfit}. In the case of robust topology optimization, we minimize the expected mean of squared differences between the target release profile and the normalized design release profile. 

For a given target release profile, we run four different experiments to design the composition of the drug. In the first experiment, we use a deterministic design with a distance filter having a characteristic length or radius of $d_{\text{max}}=0.15\,\textrm{mm}$ \Cref{density}. The radius filter is increased to $d_{\text{max}}= 0.30\,\textrm{mm}$ for the second experiment. In the third and fourth experiments, we apply the SROM design scheme with the same objective function but using the two radius filter parameters $d_{\text{max}}= 0.15\,\textrm{mm}$ and $d_{\text{max}}= 0.30\,\textrm{mm}$, such as in the deterministic design. 

We minimize the expected sum of squares of release profile differences in the SROM method-based numerical experiments. We do not consider the standard deviation of the sum of squares of release profile differences in the SROM objective function. Including the standard deviation in the objective function would be trivial. Hence, we optimize the SROM designs without the standard deviation term to see how SROM changes the design shape and the release profile uncertainty. Including both the expected value and the standard deviation of the sum of squared differences of the release profiles within the objective function would lead to a multi-objective optimization problem, which is beyond our interest. Instead, our objective is to apply the SROM and compare the resulting design with the deterministic design. Two different radius values are selected in the distance filtering to investigate the impact of a larger radius on the complexity of the design and release profile of the design. We select 40 weighted SROM samples for approximating the Gamma distributions of the dissolution rates because previous studies have shown this number offers a good tradeoff \cite{torres2021robust}.

\subsection{Linear target release profile}

In one of the numerical experiments, we seek a drug composition that yields a linear release profile, meaning that the release rate of the drug is constant. The linear release profile, also known as the zero-order release, is of high interest in medical applications because a typical homogeneous drug releases the API faster at the beginning of the release and gradually decreases the release rate before completely dissolving. A constant release rate helps to maintain a stable API concentration in the patient's body \cite{linearmotivation}. The gradually decreasing release rate of a drug made of one homogeneous material stems from the basic Noyes-Whitney model. The surface-volume ratio is initially the largest but decreases during dissolution. The decreasing surface-volume ratio decreases the release rate, respectively.

We use two materials in the drug, having the dissolution rates of $v_1 = 0.003\,\textrm{mm/min}$ and $v_2 = 0.03\,\textrm{mm/min}$. The materials are assumed to have the same API concentration of $c_1 = c_2 = 1\,\textrm{mg/cm}^{3}$, so the design depends only on the release rates of the materials. If the drug is made only of the second material, the drug dissolves completely around $t=153\,\textrm{min}$. Similarly, the drug dissolves around it $t=1530\,\textrm{min}$ if it is made of the first material, which dissolves ten times slower than the second material. This means the designed drug must dissolve completely between these two extreme times. We select the end time of $t=750\,\textrm{min}$ for the linear release profile. We discretize the target linear release profile into 20 equispaced time points between $t= 0\,\textrm{min}$ and $t = 750\,\textrm{min}$.  

In the SROM design, 40 weighted samples for the dissolution rates are used in the optimization. The dissolution rates are assumed to be independently distributed. A Gamma distribution with a mean of $0.003\,\textrm{mm/min}$ and variance of $10^{-6}\,\textrm{mm}^2\textrm{/min}^2$ is used for the first material. The second material is assumed to follow a Gamma distribution with a mean of $0.03\,\textrm{mm/min}$ and a variance of $10^{-4}\,\textrm{mm}^2\textrm{/min}^2$. The analytical cumulative distribution functions of the dissolution rates and the cumulative distribution functions of the SROM approximation are plotted in \Cref{cdflinear}.

\subsubsection{Results}

\begin{figure*}[htb]
    \centering
     \begin{subfigure}[b]{\textwidth}
    \includegraphics[width=\textwidth]{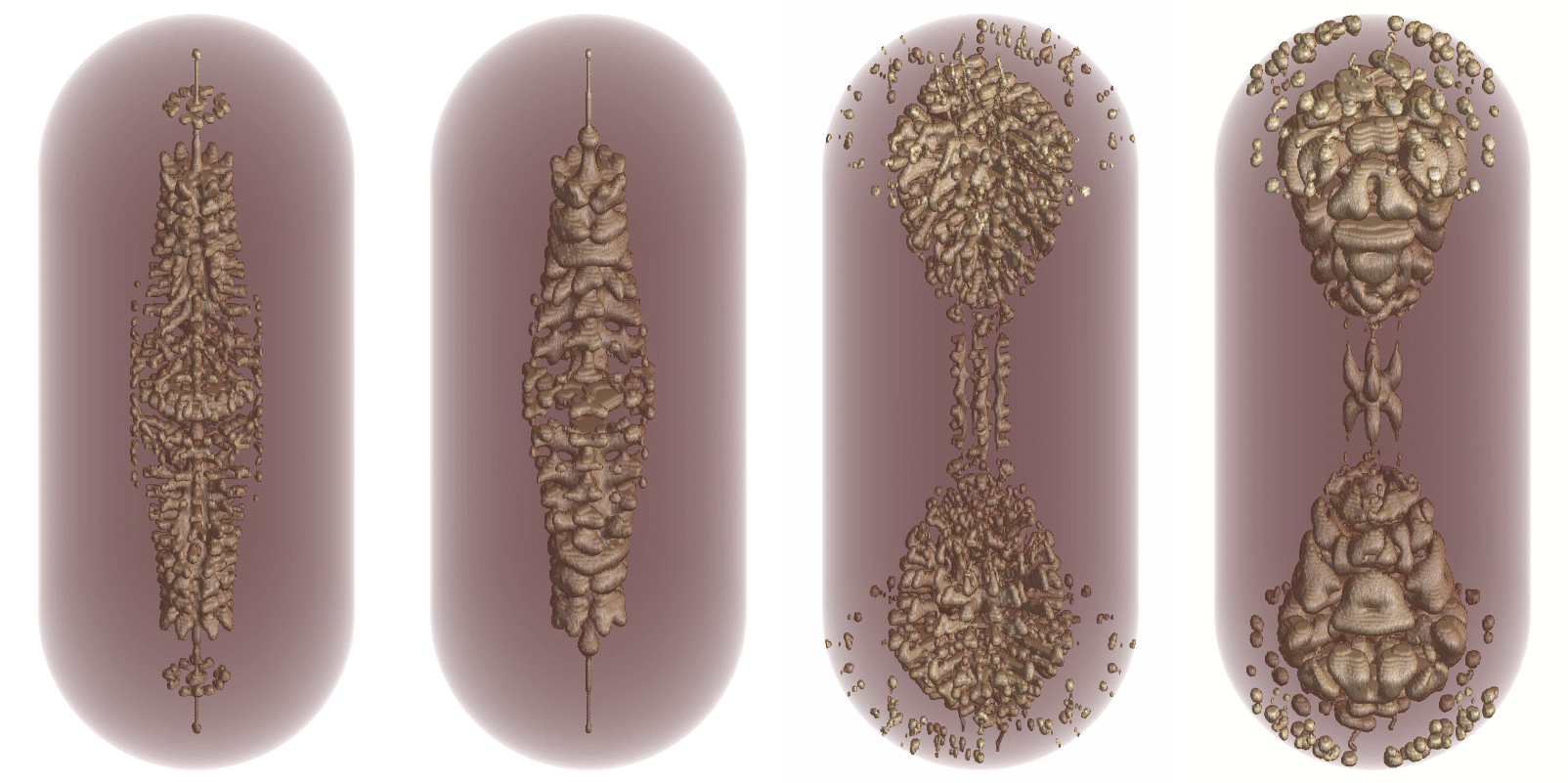}
    \caption{Four design profiles for linear release profiles. First from the left design: deterministic design with density filter with $d_{\text{max}}=0.15\,\textrm{mm}$. Second from the left design: deterministic design with density filter with $d_{\text{max}}=0.30\,\textrm{mm}$. Third from left design: SROM design with density filter with $d_{\text{max}}=0.15\,\textrm{mm}$. Fourth from the left: SROM design with density filter with $d_{\text{max}}=0.30\,\textrm{mm}$.}
         \label{linear_designs}
     \end{subfigure}

    \begin{subfigure}[b]{\textwidth}
        \centering
        \includegraphics[width=\textwidth]{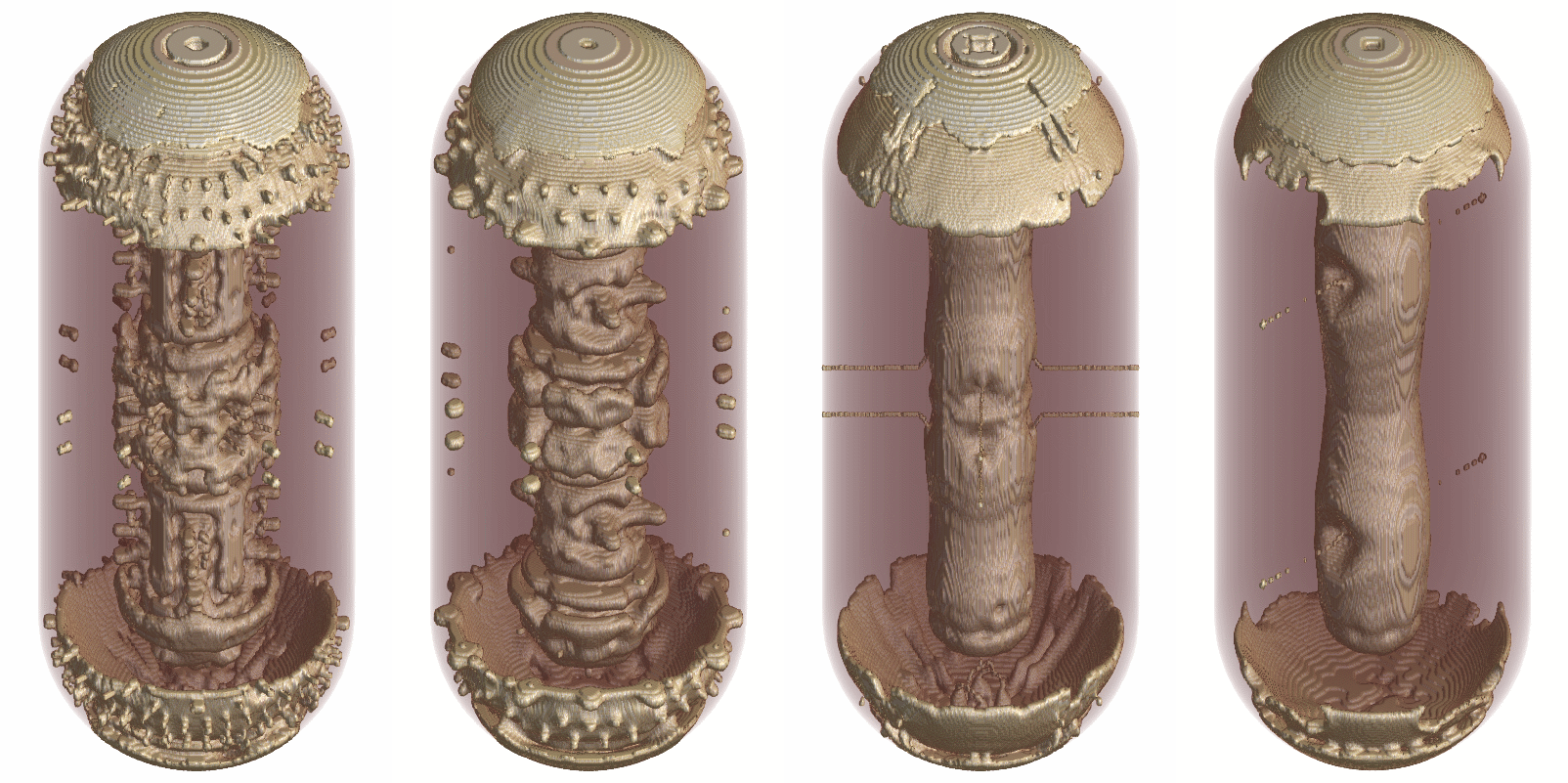}
         \caption{Four designs for the pulsatile release profile. First from the left design: deterministic design with density filter with $d_{\text{max}}=0.15\,\textrm{mm}$. Second from the left design: deterministic design with density filter with $d_{\text{max}}=0.30\,\textrm{mm}$. Third from left design: SROM design with density filter with $d_{\text{max}}=0.15\,\textrm{mm}$. Fourth from the left: SROM design with density filter with $d_{\text{max}}=0.30\,\textrm{mm}$.}
        \label{pulsatile_designs}
    \end{subfigure}
    \caption{Semitransparent volume plots of the optimal designs. The material with a dissolution rate of $v_1$ is plotted in transparent red, and the material with a rate of $v_2$ is plotted in opaque beige.}
    \Description{ The figure contains eight vertically oriented, oval-shaped drugs that are semitransparent. The capsules are rendered soft brown against a muted purple background. The background refers to the material with a slower dissolution rate, and the brown color represents the faster-dissolving material. In the first top design from the left, the material distribution of the faster-dissolving material consists of a thin, elongated structure resembling a central spine with smaller, symmetrical, branch-like extensions on both sides. The structure appears layered and vertically stacked, almost like ribs arranged along a central axis. The second top design from left: a thicker, more compact central column, still vertically aligned, with a denser texture. Symmetry is maintained, and the overall look is more solid and less branched compared to the first. The right top design, from left: a more textured and irregular formation of the faster-dissolving material, with two large masses at the top and bottom, connected by a narrower central spine.
It is surrounded by scattered small particles, giving it a diffuse, cloudy aura. The fourth top figure from the left features a highly detailed, biomorphic formation, where the top and bottom parts appear almost like faces or heads, facing each other. 

The following is a description of four additional plots at the bottom. The first bottom figure from the left: the structure of the faster-dissolving material resembles a twisted, highly detailed central column connecting a dome-like top and a bowl-shaped bottom. The surface of the faster-dissolving material is rough and irregular, featuring numerous small, bubble-like bumps and voids, which give it a chaotic or coral-like appearance. The second bottom figure from the left has a similar dome and bowl structure of the faster-dissolving material. Still, the central column is slightly less chaotic, with more defined ridges. The texture still exhibits complexity but is slightly smoother than the first. The third bottom figure from the left, the top dome of the faster-dissolving material distribution, is more geometrically structured, with concentric rings. The central column is much smoother and cylindrical, showing a more uniform texture. The fourth bottom figure from the left: The structure of the faster-dissolving material is very smooth and cylindrical, resembling a polished column that connects the same dome and bowl shapes. All complex surface features are removed. }
    \label{fig:combined_volumes}
\end{figure*}

The final designs of the drug compositions tailored for the linear target release profile are illustrated in \Cref{linear_designs}. The compositions obtained by the deterministic design approach have thin vertical channels made of the second material, which has a higher dissolution rate. These thin channels guide the dissolution to proceed toward the inner parts of the drug at the beginning of the release. To achieve a linear release profile with the given materials and their dissolution rates, both of the materials must dissolve simultaneously. Otherwise, the release profile would be decreasing at least part of the time non-linearly. The thin vertical channel appears to provide a mechanism for the simultaneous release of the two materials. The outer layer of the drug consists of the slower-dissolving material, with dissolution progressing from the outside inward. The drug also dissolves from the thin vertical channels and chambers connected to them, meaning that both of the materials indeed dissolve at the same time. 

\begin{figure*}[htb]
     \centering
     \includegraphics[width=\textwidth]{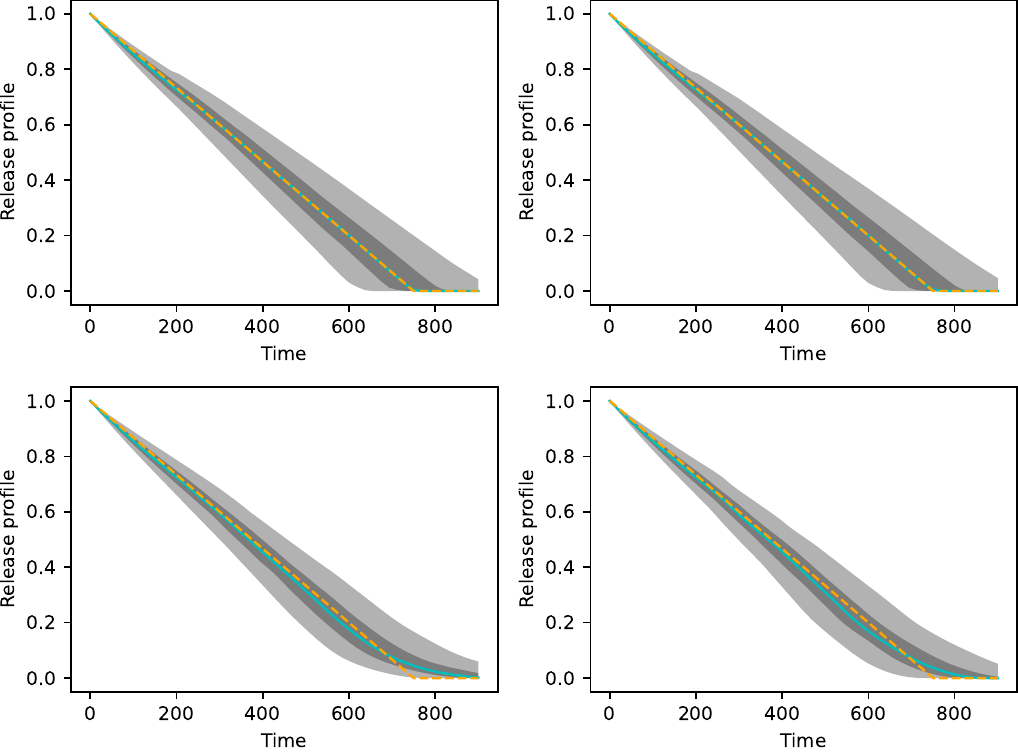}

   \caption{Distributions of release profiles of designs for linear target release. Orange curves stand for the target release profiles. The cyan curve is the release profile of the design when the dissolution rate parameters are fixed at their ideal values ($v_1 = 0.003\,\textrm{mm/min}$ and $v_2 = 0.03\,\textrm{mm/min}$). The light grey area denotes the 95th and 5th quantiles of the release distributions of the designs when the dissolution rate parameters follow a Gamma distribution. The dark grey regions represent the 75th and 25th quantiles of the release distributions, respectively. Top left plot: the release profiles of the deterministic design with density filter with $d_{\text{max}}=0.15\,\textrm{mm}$. Top right plot: the release profiles of the deterministic design with density filter with $d_{\text{max}}=0.30\,\textrm{mm}$. Bottom left plot: the release profiles of the SROM design with density filter with $d_{\text{max}}=0.15\,\textrm{mm}$. Bottom right plot: the release profiles of the SROM design with density filter with $d_{\text{max}}=0.30\,\textrm{mm}$.}
   
\Description{The figure presents four plots showing drug release profiles over time for designs that target a linear release. Each plot includes an orange line representing the target release and a cyan line showing the design profile under fixed (ideal) drug dissolution rate parameters. The shaded gray regions depict uncertainty, with light gray representing the 5th to 95th percentiles and dark gray representing the 25th to 75th percentiles of the release distributions under Gamma-distributed dissolution rate parameters. The top two plots show deterministic designs with density filter values of $d_{\text{max}} = 0.15$ (left) and $0.30$ (right). In comparison, the bottom two plots display Stochastic Reduced Order Model (SROM) designs with the same respective $d_{\text{max}}$ values. SROM designs generally show reduced variability compared to deterministic designs, especially in late time periods, indicating improved robustness under uncertainty.}
   
     \label{linear_release}
      \end{figure*}

\begin{figure*}[htb]
     \centering
     \includegraphics[width=\textwidth]{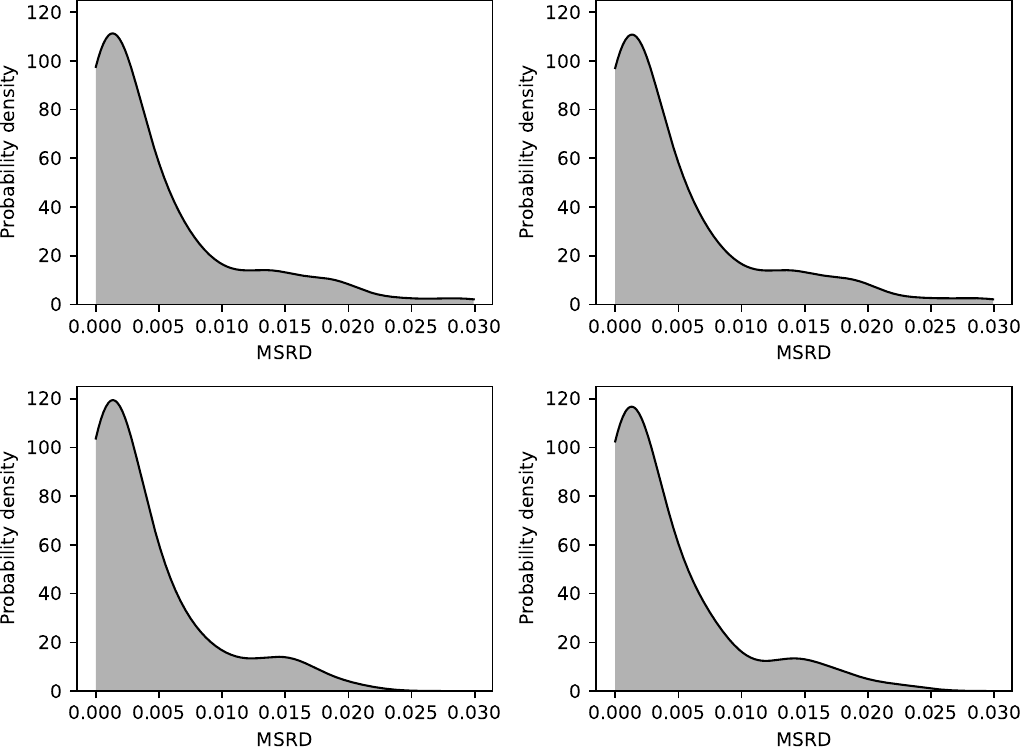}

   \caption{Kernel density estimates of the mean squared release differences (MSRD) for the linear target profiles designs when the dissolution rate parameters follow the Gamma distribution. A Gaussian function with the bandwidth of $w=0.0022$ is used as a kernel. The mean squared release difference is computed for the time interval of $t \in [0, 750]$. Top left: probability distribution of the deterministic design with density filter with $d_{\text{max}}=0.15\,\textrm{mm}$. Top right: probability distribution of the deterministic design with density filter with $d_{\text{max}}=0.30\,\textrm{mm}$. Bottom left: probability distribution of the SROM design with density filter with $d_{\text{max}}=0.15\,\textrm{mm}$. Bottom right: probability distribution of the SROM design with density filter with $d_{\text{max}}=0.30\,\textrm{mm}$.}
   \Description{This figure shows kernel density estimates of the mean squared release difference (MSRD) for linear drug release designs, comparing deterministic and SROM (stochastic) approaches under two density filter settings: $d_{\text{max}} = 0.15\,\textrm{mm}$ and $d_{\text{max}} = 0.30\,\textrm{mm}$. MSRD distribution of the SROM designs is slightly better in terms of reduced variability (lighter tails and smaller mean).}
     \label{linear_probability}
      \end{figure*}

The compositions obtained via the SROM approach are more scattered from their hemisphere parts than the designs computed with the deterministic method. The increased degree of scattering may be due to the fact that SROM designs lack thin vertical channels in their middle sections. If the channels are not available, the linear release profile must be acquired by other means. The disjoint and scattered pieces of the second material embedded in the first material are one way to accomplish that. When the dissolving boundary of the drug propagates inward, two materials are present simultaneously, just like in the deterministic designs. It is highly likely that the presence of the thin channels makes the release less robust. That is because if the dissolution rates are random variables, the causality of the Eikonal equation accumulates uncertainty as the dissolution proceeds. The thin channels manifest as increased uncertainty at the end of the release profiles, as shown in \Cref{linear_release}. However, the kernel density estimates of the mean squared release differences of the designs are not significantly different between the deterministic and the SROM designs. The kernel density estimates are plotted in \Cref{linear_probability}. Overall, the SROM designs gave slightly smaller expected mean squared release profile differences than the deterministic designs. Interestingly, the increased robustness increases the misfit between the target release curve and the release profile of the SROM design using the ideal dissolution rates $v_1=0.003\,\textrm{mm/min}$ and $v_2=0.003\,\textrm{mm/min}$, which is evident from the release curves in \Cref{linear_designs}. 

The distance filtering seems to have a negligible effect on the design profiles themselves. However, increasing the filtering radius makes the designs more regular and less detailed in both the deterministic and the SROM designs. This is a good result from the manufacturability perspective.

\subsection{Pulsatile target release profile}

The pulsatile target release profile is also interesting in controlled-release drug delivery systems because symptoms of certain diseases follow the circadian rhythm \cite{pulsatile}, such as specific types of asthma \cite{pulsatile2}. Synchronizing the release rate with the prevalence of symptoms is beneficial for the patient's treatment \cite{pulsatile2}. 

In the numerical experiments of the pulsatile release profiles, the dissolution rate of a first material is $v_1= 0.01\,\textrm{mm/min}$, and the API concentration is $c_1= 0.0001\,\textrm{mg/cm}^{3}$. The dissolution rate of the second material is $v_2=0.03\,\textrm{mm/min}$, and the API concentration is $c_2=1\,\textrm{mg/cm}^{3}$. Our target release profile of the remaining drug decreases linearly from $100\%$ to $60\%$ between time $t=0\,\textrm{min}$ and $t=60\,\textrm{min}$. The target release is paused between times $t=60\,\textrm{min}$ and $t=210\,\textrm{min}$. The target release continues at $t=210\,\textrm{min}$ and maintains a constant rate until $t=300\,\textrm{min}$, when the target drug is expected to be completely dissolved. The target release profile is discretized into 16 equispaced points for $t \in [0,300]$. 

\begin{figure*}[htb]
    \centering
     \begin{subfigure}[b]{\textwidth}
    \includegraphics[width=\textwidth]{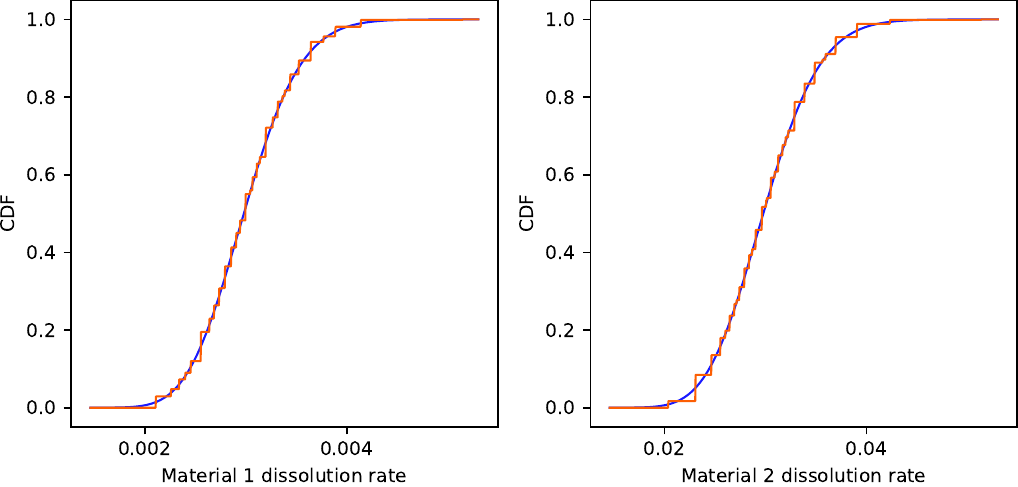}
    \caption{The cumulative distribution functions of the dissolution rates for the linear release profile.}
      \label{cdflinear}
     \end{subfigure}
   
\vspace{5mm}
    \begin{subfigure}[b]{\textwidth}
        \centering
        \includegraphics[width=\textwidth]{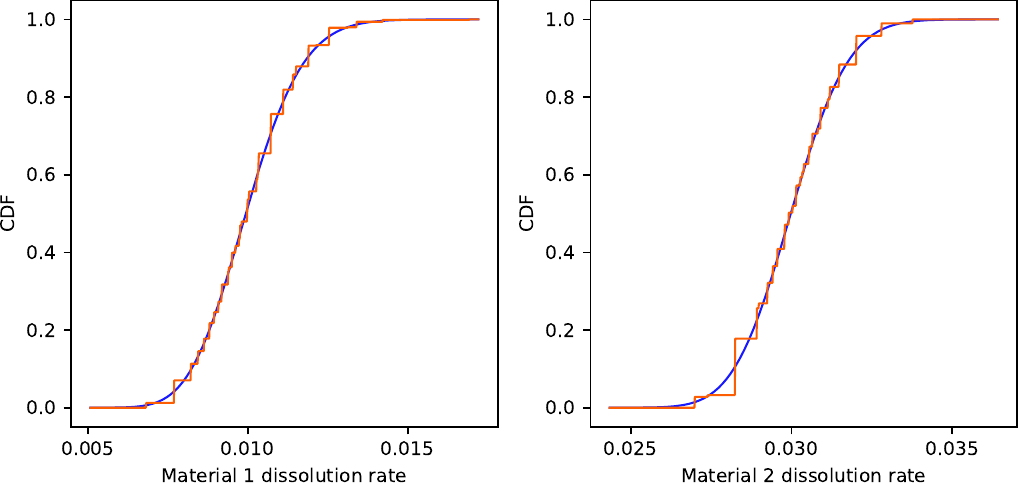}
         \caption{The cumulative distribution functions of the dissolution rates for the pulsatile release profile.}
        \label{cdfpulsa}
    \end{subfigure}
    \caption{The cumulative distribution functions of the dissolution rates and the SROM approximations of the distributions. Blue: analytical CDF of the Gamma distribution. Red: the CDF of the SROM approximation. The SROM method can accurately match the Gamma distribution of the dissolution rate with 40 weighted samples.}
    \Description{This figure shows the CDFs of dissolution rates for two materials under linear and pulsatile drug release profiles. Each plot compares the analytical Gamma CDF (blue) to the SROM approximation (red). The SROM method closely matches the true distribution across all cases, demonstrating its accuracy using only 40 weighted samples. Because the SROM is based on weighted discrete samples, its CDFs exhibit discontinuities at the sample values. This manifests in stairs-like CDF functions.}
    \label{fig:combined_subfigures}
\end{figure*}

In the SROM design, we assume the dissolution rates of the materials have less uncertainty than the linear target release dissolution rate parameters.  Both dissolution rates are assumed to follow Gamma distributions and be independently distributed. The mean of the first material dissolution rate is $0.01\,\textrm{mm/min}$, and its variance is $2\times10^{-6}\,\textrm{mm}^2\textrm{/min}^2$. The mean of the second material dissolution rate is assumed to be $0.03\,\textrm{mm/min}$, and the variance is $2\times10^{-6}\,\textrm{mm}^2\textrm{/min}^2$. 40 weighted SROM samples are used in the optimization. The cumulative distribution functions of the dissolution rate parameters and their approximations, as given by the SROM, are illustrated in \Cref{cdfpulsa}.

\subsubsection{Results} 

The volume plots of the final designs for the pulsed target release are plotted in \Cref{pulsatile_designs}. As can be seen from the volume plots, the deterministic designs have more details than the SROM designs. The deterministic designs consist of small overhangs. The surface of the second material of the deterministic designs is also much rougher than that of the SROM designs. An increased distance filtering radius $d_{\text{max}}=0.30\,\textrm{mm}$ results in slightly less detailed designs compared to the $d_{\text{max}}=0.15\,\textrm{mm}$, but partially in the case of the SROM designs, the effect is relatively small.

\begin{figure*}[htb]
     \centering
     \includegraphics[width=\textwidth]{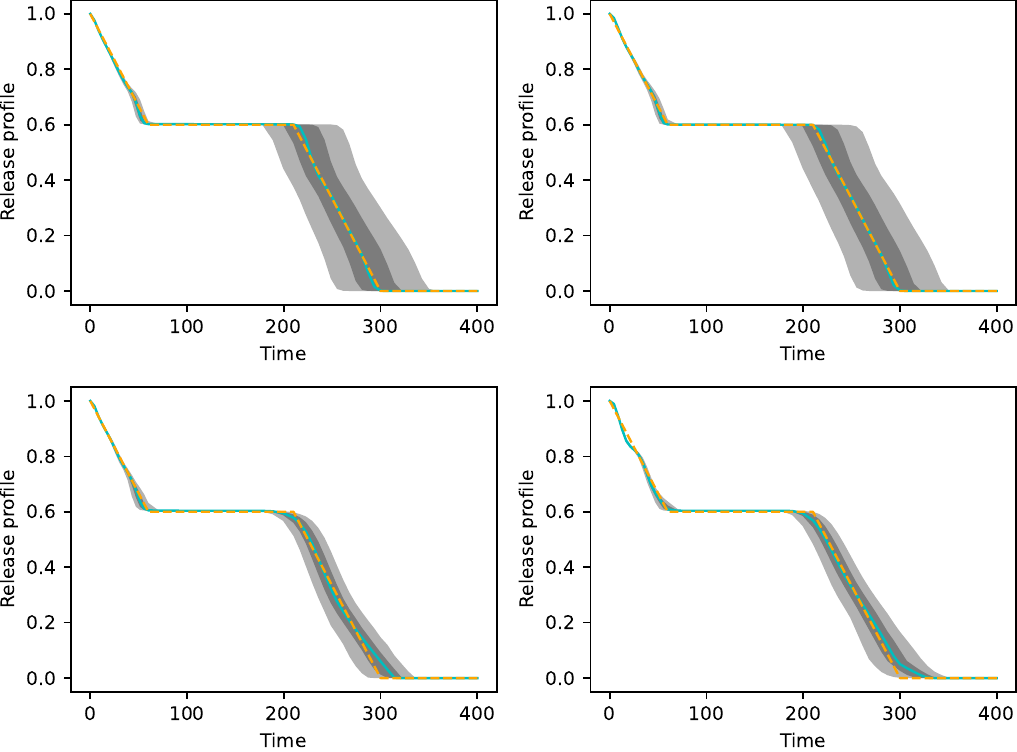}

   \caption{Distributions of release profiles of designs for pulsatile target release. Orange curves stand for the target release profiles. The cyan curve is the release profile of the design when the dissolution rate parameters are fixed at their ideal values ($v_1 = 0.01\,\textrm{mm/min}$ and $v_2 = 0.03\,\textrm{mm/min}$). The light grey area denotes the 95th and 5th quantiles of the release distributions of the designs when the dissolution rate parameters follow the Gamma distribution. The dark grey regions imply the 75th and 25th quantiles of the release distributions, respectively. Top left plot: the release profiles of the deterministic design with density filter with $d_{\text{max}}=0.15\,\textrm{mm}$. Top right plot: the release profiles of the deterministic design with density filter with $d_{\text{max}}=0.30\,\textrm{mm}$. Bottom left plot: the release profiles of the SROM design with density filter with $d_{\text{max}}=0.15\,\textrm{mm}$. Bottom right plot: the release profiles of the SROM design with density filter with $d_{\text{max}}=0.30\,\textrm{mm}$.
   }

   \Description{The figure presents four plots showing drug release profiles over time for designs that target a pulsatile release. Each plot includes an orange line representing the target release and a cyan line showing the design profile under fixed (ideal) drug dissolution rate parameters. The shaded gray regions depict uncertainty, with light gray representing the 5th to 95th percentiles and dark gray representing the 25th to 75th percentiles of the release distributions under Gamma-distributed dissolution rate parameters. The top two plots show deterministic designs with density filter values of $d_{\text{max}} = 0.15$ (left) and $0.30$ (right). In comparison, the bottom two plots display Stochastic Reduced Order Model (SROM) designs with the same respective $d_{\text{max}}$ values. SROM designs significantly reduce the uncertainty compared to deterministic designs. This effect is most visible in the release profile after the pulse time periods.}
     \label{pulsatile_release}
      \end{figure*}

The release profiles of the deterministic designs are slightly closer to the target release curve than those of the SROM designs in the best-case scenario, where the dissolution rates match their ideal exact values \ref{pulsatile_release}. The small details in the design might help to achieve the best overall agreement with the target release profile. The SROM designs have a slightly worse misfit in the best-case scenario but are not significantly inferior to the deterministic designs. On the other hand, the SROM designs are much smoother and more regular. We hypothesize that the increased regularity of SROM designs enables them to exhibit improved resilience against uncertainty. This result is reasonable, as increased tolerances typically lead to designs that are more robust to uncertainties. The SROM designs also show a significantly smaller variance in the marginal distributions of the probabilistic release curves when dissolution rates follow Gamma distributions \ref{pulsatile_release}. Due to the relatively slight variance of the dissolution rates, the probabilistic release curves are close to each other during that $t\in[0,210]$. The uncertainty of the releases for $t\in[210,300]$ is much smaller in the SROM designs than in the deterministic designs. 

\begin{figure*}[htb]
     \centering
     \includegraphics[width=\textwidth]{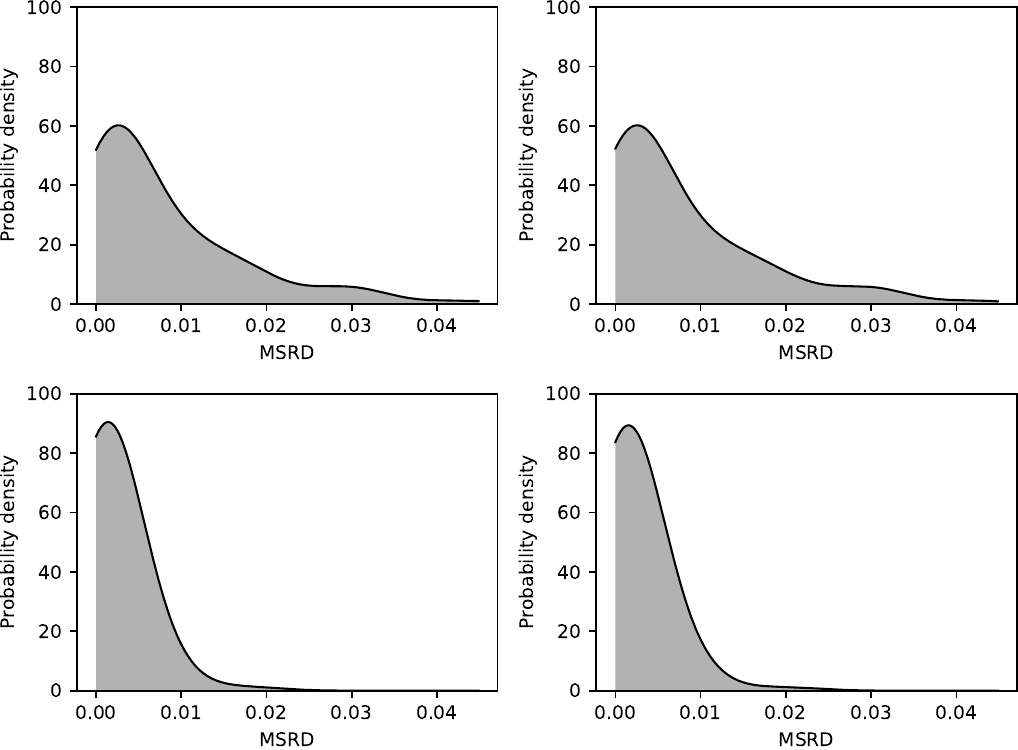}
      
   \caption{Kernel density estimates of the mean squared release differences of the designs for the pulsatile target profiles when the dissolution rate parameters follow the Gamma distribution. The mean squared release difference (MSRD) is computed for the time interval of $t \in [0, 400]$. A Gaussian function with the bandwidth of $w=0.004$ is used as the kernel. Top left: probability distribution of the deterministic design with density filter with $d_{\text{max}}=0.15\,\textrm{mm}$. Top right: probability distribution of the deterministic design with density filter with $d_{\text{max}}=0.30\,\textrm{mm}$. Bottom left: probability distribution of the SROM design with density filter with $d_{\text{max}}=0.15\,\textrm{mm}$. Bottom right: probability distribution of the SROM design with density filter with $d_{\text{max}}=0.30\,\textrm{mm}$.}
      \Description{This figure shows kernel density estimates of the mean squared release difference (MSRD) for pulsatile drug release designs, comparing deterministic and SROM (stochastic) approaches under two density filter settings: $d_{\text{max}} = 0.15$ and $d_{\text{max}} = 0.30$. SROM designs have a significantly smaller variance in their MSRD distribution. The tails of the distribution are much lighter, and the mean is also smaller. }
     \label{pulsatile_probability}
      \end{figure*}

The robustness of the SROM designs is seen best in \Cref{pulsatile_probability}, which is the plot of the distribution of the mean squared release profile differences of the designs during time $t\in[0,300]$. Not only is the mean of the expected squared release profile difference smaller, but the variance of the distribution is also substantially improved in comparison to the same probability distribution of the deterministic designs.

\FloatBarrier
\section{Conclusion and future works}
\label{conclusionsection}

We implemented a topology optimization approach for the optimal design of personalized drugs. The proposed method computes the optimal drug composition to yield a release profile that closely matches the target, offering degrees of freedom through its non-parametric principle. The method assumes that the drug dissolves from its boundary, meaning that the Noyes-Whitney dissolution model governs the release. The evolution of the dissolving drug was tracked by the Eikonal equation, which was approximated numerically by the first-order fast marching method. An accurate and numerically efficient marching cubes-based method was employed to calculate the mass of the remaining drug during the dissolution process. This eliminated the need for using heuristic approximations to evaluate the implicit volume defined by the Eikonal equation. The numerical choices also allowed for effective sensitivity analysis via the discrete adjoint method. To design the optimal material distribution, the proposed optimal design method must be supplied with the target release profile time series, the dissolution rates of the materials, and the shape of the drug. By optimizing the entire three-dimensional composition from the ground up, our approach avoids the assumption that certain parts of the drug remain undissolved.

Additionally, the method can address the uncertainty in the dissolution rates of the materials through the SROM. Compared to classical Monte Carlo methods, the SROM approach offers an efficient method for optimizing material composition under uncertainty by minimizing an objective function that depends on random dissolution rates. The proposed methods were applied to design two different release profiles. The SROM approach reduces uncertainty in the release profiles, specifically in terms of the expected squared difference in release profiles. The drug compositions also differ from those generated using the non-probabilistic approach, demonstrating the non-trivial benefits provided by the SROM method.

One possible future direction could be to generalize the dissolution model to diffusion-governed releases, which are prevalent in pharmaceutical studies. This would require significantly more numerical side work to define the adjoint equation for the gradients of the objective functions, but we expect this to be possible. Another research direction could involve incorporating the manufacturability of the drugs as an additional objective within a multi-objective optimization framework alongside the misfit functions. Although the current additive manufacturing technology is advanced, it might not be flexible enough to produce the complex and scattered designs that the promised numerical method outputs. However, manufacturability might be challenging to formalize as an objective in the optimization because our drug design consists solely of solid materials, and the final capsules do not contain voids. The manufacturability objectives might involve factors such as the steepness of material boundary angles and material connectivity. The selected additive manufacturing method also influences these constraints.

\begin{acks}
The authors thank Andreas Rupp for insightful discussions related to the mathematical aspects of the manuscript, particularly the adjoint method. The authors also thank Alaa Mahran and Rathna Mathiyalagan for the helpful discussions on understanding the pharmaceutical perspective of personalized drug release. We acknowledge funding from Business Finland (project numbers 539/31/2023, 147/31/2023) and the Research Council of Finland for the Flagship of Advanced Mathematics for Sensing, Imaging, and Modelling 2024–2031 (decision number 359183) and for the Centre of Excellence in Inverse Modelling and Imaging 2018–2025 (decision number 353095).


\end{acks}

\bibliographystyle{ACM-Reference-Format}
\bibliography{bibliography}


\begin{thebibliography}{69}


\ifx \showCODEN    \undefined \def \showCODEN     #1{\unskip}     \fi
\ifx \showISBNx    \undefined \def \showISBNx     #1{\unskip}     \fi
\ifx \showISBNxiii \undefined \def \showISBNxiii  #1{\unskip}     \fi
\ifx \showISSN     \undefined \def \showISSN      #1{\unskip}     \fi
\ifx \showLCCN     \undefined \def \showLCCN      #1{\unskip}     \fi
\ifx \shownote     \undefined \def \shownote      #1{#1}          \fi
\ifx \showarticletitle \undefined \def \showarticletitle #1{#1}   \fi
\ifx \showURL      \undefined \def \showURL       {\relax}        \fi
\providecommand\bibfield[2]{#2}
\providecommand\bibinfo[2]{#2}
\providecommand\natexlab[1]{#1}
\providecommand\showeprint[2][]{arXiv:#2}

\bibitem[Agarwal(2004)]%
        {rbdo}
\bibfield{author}{\bibinfo{person}{Harish Agarwal}.}
  \bibinfo{year}{2004}\natexlab{}.
\newblock \emph{\bibinfo{title}{{Reliability Based Design Optimization:
  Formulations and Methodologies}}}.
\newblock \bibinfo{thesistype}{Ph.\,D. Dissertation}.
  \bibinfo{school}{University of Notre Dame}.
\newblock
\href{https://doi.org/10.7274/rb68x922w8j}{doi:\nolinkurl{10.7274/rb68x922w8j}}


\bibitem[Allaire et~al\mbox{.}(2021)]%
        {allaire2021shape}
\bibfield{author}{\bibinfo{person}{Grégoire Allaire}, \bibinfo{person}{Charles
  Dapogny}, {and} \bibinfo{person}{François Jouve}.}
  \bibinfo{year}{2021}\natexlab{}.
\newblock \bibinfo{booktitle}{\emph{Chapter 1 - Shape and topology
  optimization}}. \bibinfo{series}{Handbook of Numerical Analysis},
  Vol.~\bibinfo{volume}{22}.
\newblock \bibinfo{publisher}{Elsevier}, \bibinfo{address}{Amsterdam, The
  Netherlands}, \bibinfo{pages}{1--132}.
\newblock
\showISSN{1570-8659}
\href{https://doi.org/10.1016/bs.hna.2020.10.004}{doi:\nolinkurl{10.1016/bs.hna.2020.10.004}}


\bibitem[Altunay et~al\mbox{.}(2024)]%
        {ECMS}
\bibfield{author}{\bibinfo{person}{Rabia Altunay}, \bibinfo{person}{Jarkko
  Suuronen}, \bibinfo{person}{Andreas Rupp}, \bibinfo{person}{Eero Immonen},
  {and} \bibinfo{person}{Lassi Roininen}.} \bibinfo{year}{2024}\natexlab{}.
\newblock \showarticletitle{Reinforcement approach using topology
  optimization}. In \bibinfo{booktitle}{\emph{ECMS 2024, 38th Proceedings}}.
  \bibinfo{publisher}{The European Council for Modelling and Simulation},
  \bibinfo{address}{Cracow, Poland}, \bibinfo{pages}{330--337}.
\newblock


\bibitem[Altunay et~al\mbox{.}(2025)]%
        {altunay2025denture}
\bibfield{author}{\bibinfo{person}{Rabia Altunay}, \bibinfo{person}{Kalevi
  Vesterinen}, \bibinfo{person}{Pasi Alander}, \bibinfo{person}{Eero Immonen},
  \bibinfo{person}{Andreas Rupp}, {and} \bibinfo{person}{Lassi Roininen}.}
  \bibinfo{year}{2025}\natexlab{}.
\newblock \showarticletitle{Denture reinforcement via topology optimization}.
\newblock \bibinfo{journal}{\emph{Medical Engineering \& Physics}}
  \bibinfo{volume}{135} (\bibinfo{year}{2025}), \bibinfo{pages}{104272}.
\newblock


\bibitem[Bend{\o}e and Sigmund(2003)]%
        {originalSIMP}
\bibfield{author}{\bibinfo{person}{Martin~P. Bend{\o}e} {and}
  \bibinfo{person}{Ole Sigmund}.} \bibinfo{year}{2003}\natexlab{}.
\newblock \bibinfo{booktitle}{\emph{Topology optimization: theory, methods, and
  applications}}.
\newblock \bibinfo{publisher}{Springer Science \& Business Media},
  \bibinfo{address}{Heidelberg, Germany}.
\newblock


\bibitem[Bends{\o}e(1995)]%
        {bendsoe1995optimization}
\bibfield{author}{\bibinfo{person}{Martin~P Bends{\o}e}.}
  \bibinfo{year}{1995}\natexlab{}.
\newblock \bibinfo{booktitle}{\emph{Optimization of structural topology, shape,
  and material}}. Vol.~\bibinfo{volume}{414}.
\newblock \bibinfo{publisher}{Springer}, \bibinfo{address}{Heidelberg,
  Germany}.
\newblock


\bibitem[Beyer and Sendhoff(2007)]%
        {robustreview}
\bibfield{author}{\bibinfo{person}{Hans-Georg Beyer} {and}
  \bibinfo{person}{Bernhard Sendhoff}.} \bibinfo{year}{2007}\natexlab{}.
\newblock \showarticletitle{Robust optimization – A comprehensive survey}.
\newblock \bibinfo{journal}{\emph{Computer Methods in Applied Mechanics and
  Engineering}} \bibinfo{volume}{196}, \bibinfo{number}{33}
  (\bibinfo{year}{2007}), \bibinfo{pages}{3190--3218}.
\newblock
\showISSN{0045-7825}
\href{https://doi.org/10.1016/j.cma.2007.03.003}{doi:\nolinkurl{10.1016/j.cma.2007.03.003}}


\bibitem[Bourdin(2001)]%
        {densityfilter}
\bibfield{author}{\bibinfo{person}{Blaise Bourdin}.}
  \bibinfo{year}{2001}\natexlab{}.
\newblock \showarticletitle{Filters in topology optimization}.
\newblock \bibinfo{journal}{\emph{Internat. J. Numer. Methods Engrg.}}
  \bibinfo{volume}{50}, \bibinfo{number}{9} (\bibinfo{year}{2001}),
  \bibinfo{pages}{2143--2158}.
\newblock
\href{https://doi.org/10.1002/nme.116}{doi:\nolinkurl{10.1002/nme.116}}


\bibitem[Cabrera et~al\mbox{.}(2006)]%
        {disso-diff}
\bibfield{author}{\bibinfo{person}{Mar{\'\i}a~I Cabrera},
  \bibinfo{person}{Julio~A Luna}, {and} \bibinfo{person}{Ricardo~JA Grau}.}
  \bibinfo{year}{2006}\natexlab{}.
\newblock \showarticletitle{Modeling of dissolution-diffusion controlled drug
  release from planar polymeric systems with finite dissolution rate and
  arbitrary drug loading}.
\newblock \bibinfo{journal}{\emph{Journal of Membrane Science}}
  \bibinfo{volume}{280}, \bibinfo{number}{1-2} (\bibinfo{year}{2006}),
  \bibinfo{pages}{693--704}.
\newblock


\bibitem[{\c{C}}omo{\u{g}}lu(2022)]%
        {kinetics}
\bibfield{author}{\bibinfo{person}{Tansel {\c{C}}omo{\u{g}}lu}.}
  \bibinfo{year}{2022}\natexlab{}.
\newblock \showarticletitle{Evaluation of Drug Release Kinetics of Temozolomide
  Loaded Plga Nanoparticles in Pluronic? F-127 Hydrogel}.
\newblock \bibinfo{journal}{\emph{Bezmialem Science}} \bibinfo{volume}{10},
  \bibinfo{number}{6} (\bibinfo{year}{2022}), \bibinfo{pages}{735--741}.
\newblock


\bibitem[Desquilbet(2022)]%
        {FMMvsFSM}
\bibfield{author}{\bibinfo{person}{Fran{\c{c}}ois Desquilbet}.}
  \bibinfo{year}{2022}\natexlab{}.
\newblock \emph{\bibinfo{title}{Fast Marching method for the computation of
  first-arrival travel time of seismic waves in anisotropic media}}.
\newblock \bibinfo{thesistype}{Ph.\,D. Dissertation}.
  \bibinfo{school}{{Universit{\'e} Grenoble Alpes}}.
\newblock


\bibitem[Dokoumetzidis et~al\mbox{.}(2006)]%
        {Noyes-Whitney}
\bibfield{author}{\bibinfo{person}{Aristides Dokoumetzidis},
  \bibinfo{person}{Vasiliki Papadopoulou}, {and} \bibinfo{person}{Panos
  Macheras}.} \bibinfo{year}{2006}\natexlab{}.
\newblock \showarticletitle{Analysis of dissolution data using modified
  versions of Noyes--Whitney equation and the Weibull function}.
\newblock \bibinfo{journal}{\emph{Pharmaceutical Research}}
  \bibinfo{volume}{23} (\bibinfo{year}{2006}), \bibinfo{pages}{256--261}.
\newblock


\bibitem[Enevoldsen and Sørensen(1994)]%
        {formsorm}
\bibfield{author}{\bibinfo{person}{Ib Enevoldsen} {and} \bibinfo{person}{John~D
  Sørensen}.} \bibinfo{year}{1994}\natexlab{}.
\newblock \showarticletitle{Reliability-based optimization in structural
  engineering}.
\newblock \bibinfo{journal}{\emph{Structural Safety}} \bibinfo{volume}{15},
  \bibinfo{number}{3} (\bibinfo{year}{1994}), \bibinfo{pages}{169--196}.
\newblock
\showISSN{0167-4730}
\href{https://doi.org/10.1016/0167-4730(94)90039-6}{doi:\nolinkurl{10.1016/0167-4730(94)90039-6}}


\bibitem[Ferrari et~al\mbox{.}(2021)]%
        {REF01}
\bibfield{author}{\bibinfo{person}{Gustavo Ferrari}, \bibinfo{person}{Arthur~T
  Mello}, \bibinfo{person}{Gilberto Melo}, \bibinfo{person}{Carlos~R {de Mello
  Roesler}}, \bibinfo{person}{Gean~V Salmoria}, \bibinfo{person}{Luiz~P {de
  Souza Pinto}}, {and} \bibinfo{person}{Izabelle {de Mello Gindri}}.}
  \bibinfo{year}{2021}\natexlab{}.
\newblock \showarticletitle{Polymeric implants with drug-releasing
  capabilities: a mapping review of laboratory research}.
\newblock \bibinfo{journal}{\emph{Drug Development and Industrial Pharmacy}}
  \bibinfo{volume}{47}, \bibinfo{number}{10} (\bibinfo{year}{2021}),
  \bibinfo{pages}{1535--1545}.
\newblock


\bibitem[Field et~al\mbox{.}(2015)]%
        {sromcompa}
\bibfield{author}{\bibinfo{person}{Richard Field}, \bibinfo{person}{Mircea
  Grigoriu}, {and} \bibinfo{person}{John Emery}.}
  \bibinfo{year}{2015}\natexlab{}.
\newblock \showarticletitle{On the efficacy of stochastic collocation,
  stochastic Galerkin, and stochastic reduced order models for solving
  stochastic problems}.
\newblock \bibinfo{journal}{\emph{Probabilistic Engineering Mechanics}}
  \bibinfo{volume}{41} (\bibinfo{year}{2015}), \bibinfo{pages}{60--72}.
\newblock
\showISSN{0266-8920}
\href{https://doi.org/10.1016/j.probengmech.2015.05.002}{doi:\nolinkurl{10.1016/j.probengmech.2015.05.002}}


\bibitem[Gasser and Schu{\"e}ller(1997)]%
        {rbrdo}
\bibfield{author}{\bibinfo{person}{Markus Gasser} {and}
  \bibinfo{person}{Gerhart~Iwo Schu{\"e}ller}.}
  \bibinfo{year}{1997}\natexlab{}.
\newblock \showarticletitle{Reliability-Based Optimization of Structural
  Systems}.
\newblock \bibinfo{journal}{\emph{Mathematical Methods of Operations Research}}
  \bibinfo{volume}{46}, \bibinfo{number}{3} (\bibinfo{year}{1997}),
  \bibinfo{pages}{287--307}.
\newblock
\showISSN{1432-5217}
\href{https://doi.org/10.1007/BF01194858}{doi:\nolinkurl{10.1007/BF01194858}}


\bibitem[Ghafouri et~al\mbox{.}(2023)]%
        {Jari2023}
\bibfield{author}{\bibinfo{person}{Mehran Ghafouri}, \bibinfo{person}{Mohsen
  Amraei}, \bibinfo{person}{Aditya Gopaluni}, \bibinfo{person}{Heidi Piili},
  \bibinfo{person}{Timo Bj{\"o}rk}, {and} \bibinfo{person}{Jari
  H{\"a}m{\"a}l{\"a}inen}.} \bibinfo{year}{2023}\natexlab{}.
\newblock \bibinfo{booktitle}{\emph{Simulation and Its Use in Additive
  Manufacturing}}.
\newblock \bibinfo{publisher}{Springer International Publishing},
  \bibinfo{address}{Cham}, \bibinfo{pages}{111--126}.
\newblock
\href{https://doi.org/10.1007/978-3-031-29082-4_6}{doi:\nolinkurl{10.1007/978-3-031-29082-4_6}}


\bibitem[Giles and Pierce(2000)]%
        {adjoexample}
\bibfield{author}{\bibinfo{person}{Michael Giles} {and} \bibinfo{person}{Niles
  Pierce}.} \bibinfo{year}{2000}\natexlab{}.
\newblock \showarticletitle{An Introduction to the Adjoint Approach to Design}.
\newblock \bibinfo{journal}{\emph{Flow, Turbulence and Combustion}}
  \bibinfo{volume}{65} (\bibinfo{date}{04} \bibinfo{year}{2000}).
\newblock
\href{https://doi.org/10.1023/A:1011430410075}{doi:\nolinkurl{10.1023/A:1011430410075}}


\bibitem[Gomes-Filho et~al\mbox{.}(2022)]%
        {diff-ero}
\bibfield{author}{\bibinfo{person}{M{\'a}rcio~S Gomes-Filho},
  \bibinfo{person}{Fernando~A Oliveira}, {and} \bibinfo{person}{Marco~AA
  Barbosa}.} \bibinfo{year}{2022}\natexlab{}.
\newblock \showarticletitle{Modeling the diffusion-erosion crossover dynamics
  in drug release}.
\newblock \bibinfo{journal}{\emph{Physical Review E}} \bibinfo{volume}{105},
  \bibinfo{number}{4} (\bibinfo{year}{2022}), \bibinfo{pages}{044110}.
\newblock


\bibitem[González-Acuña and Chaparro-Romo(2020)]%
        {eikonalinfo2}
\bibfield{author}{\bibinfo{person}{Rafael~G González-Acuña} {and}
  \bibinfo{person}{Héctor~A Chaparro-Romo}.} \bibinfo{year}{2020}\natexlab{}.
\newblock \showarticletitle{The Eikonal equation}.
\newblock In \bibinfo{booktitle}{\emph{Stigmatic Optics}}.
  \bibinfo{publisher}{IOP Publishing}, \bibinfo{address}{Bristol, UK},
  \bibinfo{pages}{2--1 to 2--13}.
\newblock
\showISBNx{978-0-7503-3463-1}
\href{https://doi.org/10.1088/978-0-7503-3463-1ch2}{doi:\nolinkurl{10.1088/978-0-7503-3463-1ch2}}


\bibitem[Grigoriu(2009)]%
        {SROMgrigoriu}
\bibfield{author}{\bibinfo{person}{Mircea Grigoriu}.}
  \bibinfo{year}{2009}\natexlab{}.
\newblock \showarticletitle{Reduced order models for random functions.
  Application to stochastic problems}.
\newblock \bibinfo{journal}{\emph{Applied Mathematical Modelling}}
  \bibinfo{volume}{33}, \bibinfo{number}{1} (\bibinfo{year}{2009}),
  \bibinfo{pages}{161--175}.
\newblock


\bibitem[Grof and {\v{S}}t{\v{e}}p{\'a}nek(2021)]%
        {parametriccekrepublic}
\bibfield{author}{\bibinfo{person}{Zden{\v{e}}k Grof} {and}
  \bibinfo{person}{Franti{\v{s}}ek {\v{S}}t{\v{e}}p{\'a}nek}.}
  \bibinfo{year}{2021}\natexlab{}.
\newblock \showarticletitle{Artificial intelligence based design of 3D-printed
  tablets for personalised medicine}.
\newblock \bibinfo{journal}{\emph{Computers \& Chemical Engineering}}
  \bibinfo{volume}{154} (\bibinfo{year}{2021}), \bibinfo{pages}{107492}.
\newblock


\bibitem[Haertel(2018)]%
        {DTU-thesis}
\bibfield{author}{\bibinfo{person}{Jan Hendrik~Klaas Haertel}.}
  \bibinfo{year}{2018}\natexlab{}.
\newblock \emph{\bibinfo{title}{Design of thermal systems using topology
  optimization}}.
\newblock \bibinfo{thesistype}{Ph.\,D. Dissertation}.
  \bibinfo{school}{Technical University of Denmark}.
\newblock


\bibitem[Hassouna and Farag(2007)]%
        {multistencilsFMM}
\bibfield{author}{\bibinfo{person}{Mohamed~S Hassouna} {and}
  \bibinfo{person}{Aly~A Farag}.} \bibinfo{year}{2007}\natexlab{}.
\newblock \showarticletitle{Multistencils fast marching methods: A highly
  accurate solution to the eikonal equation on cartesian domains}.
\newblock \bibinfo{journal}{\emph{IEEE Transactions on Pattern Analysis and
  Machine Intelligence}} \bibinfo{volume}{29}, \bibinfo{number}{9}
  (\bibinfo{year}{2007}), \bibinfo{pages}{1563--1574}.
\newblock


\bibitem[Huang and Arora(1997)]%
        {discrete}
\bibfield{author}{\bibinfo{person}{Min-Wei Huang} {and}
  \bibinfo{person}{Jasbir~S Arora}.} \bibinfo{year}{1997}\natexlab{}.
\newblock \showarticletitle{Optimal design with discrete variables: some
  numerical experiments}.
\newblock \bibinfo{journal}{\emph{Internat. J. Numer. Methods Engrg.}}
  \bibinfo{volume}{40}, \bibinfo{number}{1} (\bibinfo{year}{1997}),
  \bibinfo{pages}{165--188}.
\newblock


\bibitem[Iyer et~al\mbox{.}(2006)]%
        {ref0}
\bibfield{author}{\bibinfo{person}{Sunil~S Iyer}, \bibinfo{person}{William~H
  Barr}, {and} \bibinfo{person}{Henry~T Karnes}.}
  \bibinfo{year}{2006}\natexlab{}.
\newblock \showarticletitle{Profiling in vitro drug release from subcutaneous
  implants: a review of current status and potential implications on drug
  product development}.
\newblock \bibinfo{journal}{\emph{Biopharmaceutics \& Drug Disposition}}
  \bibinfo{volume}{27}, \bibinfo{number}{4} (\bibinfo{year}{2006}),
  \bibinfo{pages}{157--170}.
\newblock


\bibitem[Komini et~al\mbox{.}(2023)]%
        {robustexample}
\bibfield{author}{\bibinfo{person}{Ludian Komini}, \bibinfo{person}{Matthijs
  Langelaar}, {and} \bibinfo{person}{Benedikt Kriegesmann}.}
  \bibinfo{year}{2023}\natexlab{}.
\newblock \showarticletitle{Robust topology optimization considering part
  distortion and process variability in additive manufacturing}.
\newblock \bibinfo{journal}{\emph{Advances in Engineering Software}}
  \bibinfo{volume}{186} (\bibinfo{year}{2023}), \bibinfo{pages}{103551}.
\newblock
\showISSN{0965-9978}
\href{https://doi.org/10.1016/j.advengsoft.2023.103551}{doi:\nolinkurl{10.1016/j.advengsoft.2023.103551}}


\bibitem[Laaksonen et~al\mbox{.}(2009)]%
        {cellular}
\bibfield{author}{\bibinfo{person}{Timo~J Laaksonen}, \bibinfo{person}{Hannu~M
  Laaksonen}, \bibinfo{person}{Jouni~T Hirvonen}, {and} \bibinfo{person}{Lasse
  Murtom{\"a}ki}.} \bibinfo{year}{2009}\natexlab{}.
\newblock \showarticletitle{Cellular automata model for drug release from
  binary matrix and reservoir polymeric devices}.
\newblock \bibinfo{journal}{\emph{Biomaterials}} \bibinfo{volume}{30},
  \bibinfo{number}{10} (\bibinfo{year}{2009}), \bibinfo{pages}{1978--1987}.
\newblock


\bibitem[Laracuente et~al\mbox{.}(2020)]%
        {linearmotivation}
\bibfield{author}{\bibinfo{person}{Mei-Li Laracuente},
  \bibinfo{person}{Marina~H. Yu}, {and} \bibinfo{person}{Kevin~J. McHugh}.}
  \bibinfo{year}{2020}\natexlab{}.
\newblock \showarticletitle{Zero-order drug delivery: State of the art and
  future prospects}.
\newblock \bibinfo{journal}{\emph{Journal of Controlled Release}}
  \bibinfo{volume}{327} (\bibinfo{year}{2020}), \bibinfo{pages}{834--856}.
\newblock
\showISSN{0168-3659}
\href{https://doi.org/10.1016/j.jconrel.2020.09.020}{doi:\nolinkurl{10.1016/j.jconrel.2020.09.020}}


\bibitem[Lee and Yeo(2015)]%
        {controlled}
\bibfield{author}{\bibinfo{person}{Jinhyun~H Lee} {and} \bibinfo{person}{Yoon
  Yeo}.} \bibinfo{year}{2015}\natexlab{}.
\newblock \showarticletitle{Controlled drug release from pharmaceutical
  nanocarriers}.
\newblock \bibinfo{journal}{\emph{Chemical engineering science}}
  \bibinfo{volume}{125} (\bibinfo{year}{2015}), \bibinfo{pages}{75--84}.
\newblock


\bibitem[Levi and Schibler(2007)]%
        {pulsatile}
\bibfield{author}{\bibinfo{person}{Francis Levi} {and} \bibinfo{person}{Ueli
  Schibler}.} \bibinfo{year}{2007}\natexlab{}.
\newblock \showarticletitle{Circadian Rhythms: Mechanisms and Therapeutic
  Implications}.
\newblock \bibinfo{journal}{\emph{Annual Review of Pharmacology and
  Toxicology}} \bibinfo{volume}{47}, \bibinfo{number}{Volume 47, 2007}
  (\bibinfo{year}{2007}), \bibinfo{pages}{593--628}.
\newblock
\showISSN{1545-4304}
\href{https://doi.org/10.1146/annurev.pharmtox.47.120505.105208}{doi:\nolinkurl{10.1146/annurev.pharmtox.47.120505.105208}}


\bibitem[Li et~al\mbox{.}(2020)]%
        {eikonalinfo}
\bibfield{author}{\bibinfo{person}{Jun Li}, \bibinfo{person}{Hui Li},
  \bibinfo{person}{Hui Chen}, \bibinfo{person}{Jinrong Su},
  \bibinfo{person}{Yongsheng Liu}, {and} \bibinfo{person}{Ping Tong}.}
  \bibinfo{year}{2020}\natexlab{}.
\newblock \showarticletitle{Eikonal Equation-Based Seismic Tomography of the
  Source Areas of the 2008 Mw 7.9 Wenchuan Earthquake and the 2013 Mw 6.6
  Lushan Earthquake}.
\newblock \bibinfo{journal}{\emph{Bulletin of the Seismological Society of
  America}}  \bibinfo{volume}{110} (\bibinfo{date}{04} \bibinfo{year}{2020}),
  \bibinfo{pages}{886--897}.
\newblock
\href{https://doi.org/10.1785/0120190134}{doi:\nolinkurl{10.1785/0120190134}}


\bibitem[Li and Khandelwal(2015)]%
        {continuationscheme}
\bibfield{author}{\bibinfo{person}{Lei Li} {and} \bibinfo{person}{Kapil
  Khandelwal}.} \bibinfo{year}{2015}\natexlab{}.
\newblock \showarticletitle{Volume preserving projection filters and
  continuation methods in topology optimization}.
\newblock \bibinfo{journal}{\emph{Engineering Structures}}
  \bibinfo{volume}{85} (\bibinfo{year}{2015}), \bibinfo{pages}{144--161}.
\newblock
\showISSN{0141-0296}
\href{https://doi.org/10.1016/j.engstruct.2014.10.052}{doi:\nolinkurl{10.1016/j.engstruct.2014.10.052}}


\bibitem[Lu and Anseth(1999)]%
        {multilayers}
\bibfield{author}{\bibinfo{person}{Sanxiu Lu} {and} \bibinfo{person}{Kristi~S
  Anseth}.} \bibinfo{year}{1999}\natexlab{}.
\newblock \showarticletitle{Photopolymerization of multilaminated poly(HEMA)
  hydrogels for controlled release}.
\newblock \bibinfo{journal}{\emph{Journal of Controlled Release}}
  \bibinfo{volume}{57}, \bibinfo{number}{3} (\bibinfo{year}{1999}),
  \bibinfo{pages}{291--300}.
\newblock


\bibitem[Mahran et~al\mbox{.}(2023)]%
        {mahran}
\bibfield{author}{\bibinfo{person}{Alaa Mahran}, \bibinfo{person}{Ezgi
  {\"O}zliseli}, \bibinfo{person}{Qingbo Wang}, \bibinfo{person}{Ilayda
  {\"O}zliseli}, \bibinfo{person}{Rajendra Bhadane}, \bibinfo{person}{Chunlin
  Xu}, \bibinfo{person}{Xiaoju Wang}, {and} \bibinfo{person}{Jessica~M
  Rosenholm}.} \bibinfo{year}{2023}\natexlab{}.
\newblock \showarticletitle{Semi-solid 3D printing of mesoporous silica
  nanoparticle-incorporated xeno-free nanomaterial hydrogels for protein
  delivery}.
\newblock \bibinfo{journal}{\emph{Nano Select}} \bibinfo{volume}{4},
  \bibinfo{number}{11-12} (\bibinfo{year}{2023}), \bibinfo{pages}{598--614}.
\newblock


\bibitem[Mathiyalagan et~al\mbox{.}(2023)]%
        {personalizing}
\bibfield{author}{\bibinfo{person}{Rathna Mathiyalagan}, \bibinfo{person}{Erica
  Sj{\"o}holm}, \bibinfo{person}{Sajana Manandhar}, \bibinfo{person}{Satu
  Lakio}, \bibinfo{person}{Jessica~M Rosenholm}, \bibinfo{person}{Martti
  Kaasalainen}, \bibinfo{person}{Xiaoju Wang}, {and} \bibinfo{person}{Niklas
  Sandler}.} \bibinfo{year}{2023}\natexlab{}.
\newblock \showarticletitle{Personalizing oral delivery of nanoformed piroxicam
  by semi-solid extrusion 3D printing}.
\newblock \bibinfo{journal}{\emph{European Journal of Pharmaceutical Sciences}}
   \bibinfo{volume}{188} (\bibinfo{year}{2023}), \bibinfo{pages}{106497}.
\newblock


\bibitem[Mathiyalagan et~al\mbox{.}(2025)]%
        {rathna}
\bibfield{author}{\bibinfo{person}{Rathna Mathiyalagan}, \bibinfo{person}{Max
  Westerlund}, \bibinfo{person}{Alaa Mahran}, \bibinfo{person}{Rabia Altunay},
  \bibinfo{person}{Jarkko Suuronen}, \bibinfo{person}{Mirja Palo},
  \bibinfo{person}{Johan~O Nyman}, \bibinfo{person}{Eero Immonen},
  \bibinfo{person}{Jessica~M Rosenholm}, \bibinfo{person}{Erica Monaco},
  {et~al\mbox{.}}} \bibinfo{year}{2025}\natexlab{}.
\newblock \showarticletitle{3D printing of tailored veterinary dual-release
  tablets: a semi-solid extrusion approach for metoclopramide}.
\newblock \bibinfo{journal}{\emph{RSC Pharmaceutics}} \bibinfo{volume}{2},
  \bibinfo{number}{2} (\bibinfo{year}{2025}), \bibinfo{pages}{413--426}.
\newblock


\bibitem[Milani~Fard and Milani~Fard(2022)]%
        {modeling}
\bibfield{author}{\bibinfo{person}{Maryam Milani~Fard} {and}
  \bibinfo{person}{Amir~M Milani~Fard}.} \bibinfo{year}{2022}\natexlab{}.
\newblock \showarticletitle{Modeling drug release}.
\newblock \bibinfo{journal}{\emph{Eurasian Journal of Science and Technology}}
  \bibinfo{volume}{2}, \bibinfo{number}{1} (\bibinfo{year}{2022}),
  \bibinfo{pages}{14--33}.
\newblock


\bibitem[Narasimhan(2001)]%
        {dissolution}
\bibfield{author}{\bibinfo{person}{Balaji Narasimhan}.}
  \bibinfo{year}{2001}\natexlab{}.
\newblock \showarticletitle{Mathematical models describing polymer dissolution:
  consequences for drug delivery}.
\newblock \bibinfo{journal}{\emph{Advanced Drug Delivery Reviews}}
  \bibinfo{volume}{48}, \bibinfo{number}{2-3} (\bibinfo{year}{2001}),
  \bibinfo{pages}{195--210}.
\newblock


\bibitem[Niu et~al\mbox{.}(2011)]%
        {stiffness}
\bibfield{author}{\bibinfo{person}{Fei Niu}, \bibinfo{person}{Shengli Xu},
  {and} \bibinfo{person}{Gengdong Cheng}.} \bibinfo{year}{2011}\natexlab{}.
\newblock \showarticletitle{A general formulation of structural topology
  optimization for maximizing structural stiffness}.
\newblock \bibinfo{journal}{\emph{Structural and Multidisciplinary
  Optimization}}  \bibinfo{volume}{43} (\bibinfo{year}{2011}),
  \bibinfo{pages}{561--572}.
\newblock


\bibitem[Noyes and Whitney(1897)]%
        {Noyes-Whitney-model}
\bibfield{author}{\bibinfo{person}{Arthur~A Noyes} {and}
  \bibinfo{person}{Willis~R Whitney}.} \bibinfo{year}{1897}\natexlab{}.
\newblock \showarticletitle{The rate of solution of solid substances in their
  own solutions.}
\newblock \bibinfo{journal}{\emph{Journal of the American Chemical Society}}
  \bibinfo{volume}{19}, \bibinfo{number}{12} (\bibinfo{year}{1897}),
  \bibinfo{pages}{930--934}.
\newblock


\bibitem[{\"O}zliseli et~al\mbox{.}(2023)]%
        {ozliseli}
\bibfield{author}{\bibinfo{person}{Ezgi {\"O}zliseli}, \bibinfo{person}{Sami
  {\c{S}}anl{\i}da{\u{g}}}, \bibinfo{person}{Behice S{\"u}ren},
  \bibinfo{person}{Alaa Mahran}, \bibinfo{person}{Marjaana Parikainen},
  \bibinfo{person}{Cecilia Sahlgren}, {and} \bibinfo{person}{Jessica~M
  Rosenholm}.} \bibinfo{year}{2023}\natexlab{}.
\newblock \showarticletitle{Directing cellular responses in a nanocomposite 3D
  matrix for tissue regeneration with nanoparticle-mediated drug delivery}.
\newblock \bibinfo{journal}{\emph{Materials Today Bio}}  \bibinfo{volume}{23}
  (\bibinfo{year}{2023}), \bibinfo{pages}{100865}.
\newblock


\bibitem[Paarakh et~al\mbox{.}(2018)]%
        {non-fiction}
\bibfield{author}{\bibinfo{person}{Padmaa Paarakh},
  \bibinfo{person}{Preethy~Ani Jose}, \bibinfo{person}{Chitrali~M Setty}, {and}
  \bibinfo{person}{Peter Christoper}.} \bibinfo{year}{2018}\natexlab{}.
\newblock \showarticletitle{Release kinetics--concepts and applications}.
\newblock \bibinfo{journal}{\emph{International Journal of Pharmacy Research \&
  Technology (IJPRT)}} \bibinfo{volume}{8}, \bibinfo{number}{1}
  (\bibinfo{year}{2018}), \bibinfo{pages}{12--20}.
\newblock


\bibitem[Panetta et~al\mbox{.}(2022)]%
        {shapeadjoint}
\bibfield{author}{\bibinfo{person}{Julian Panetta}, \bibinfo{person}{Haleh
  Mohammadian}, \bibinfo{person}{Emiliano Luci}, {and} \bibinfo{person}{Vahid
  Babaei}.} \bibinfo{year}{2022}\natexlab{}.
\newblock \showarticletitle{Shape from release: Inverse design and fabrication
  of controlled release structures}.
\newblock \bibinfo{journal}{\emph{ACM Transactions on Graphics (TOG)}}
  \bibinfo{volume}{41}, \bibinfo{number}{6} (\bibinfo{year}{2022}),
  \bibinfo{pages}{1--14}.
\newblock


\bibitem[Patel et~al\mbox{.}(2021)]%
        {pulsatile2}
\bibfield{author}{\bibinfo{person}{Shriya~K. Patel}, \bibinfo{person}{Mouhamad
  Khoder}, \bibinfo{person}{Matthew Peak}, {and} \bibinfo{person}{Mohamed~A.
  Alhnan}.} \bibinfo{year}{2021}\natexlab{}.
\newblock \showarticletitle{Controlling drug release with additive
  manufacturing-based solutions}.
\newblock \bibinfo{journal}{\emph{Advanced Drug Delivery Reviews}}
  \bibinfo{volume}{174} (\bibinfo{year}{2021}), \bibinfo{pages}{369--386}.
\newblock
\showISSN{0169-409X}
\href{https://doi.org/10.1016/j.addr.2021.04.020}{doi:\nolinkurl{10.1016/j.addr.2021.04.020}}


\bibitem[Rickett and Fomel(1999)]%
        {secondorderFMM}
\bibfield{author}{\bibinfo{person}{James Rickett} {and} \bibinfo{person}{Sergey
  Fomel}.} \bibinfo{year}{1999}\natexlab{}.
\newblock \showarticletitle{A second-order fast marching eikonal solver}.
\newblock \bibinfo{journal}{\emph{Stanford Exploration Project Report}}
  \bibinfo{volume}{100} (\bibinfo{year}{1999}), \bibinfo{pages}{287--293}.
\newblock


\bibitem[Schuëller and Jensen(2008)]%
        {robustreview2}
\bibfield{author}{\bibinfo{person}{Gerhart~I Schuëller} {and}
  \bibinfo{person}{Hector~A Jensen}.} \bibinfo{year}{2008}\natexlab{}.
\newblock \showarticletitle{Computational methods in optimization considering
  uncertainties – An overview}.
\newblock \bibinfo{journal}{\emph{Computer Methods in Applied Mechanics and
  Engineering}} \bibinfo{volume}{198}, \bibinfo{number}{1}
  (\bibinfo{year}{2008}), \bibinfo{pages}{2--13}.
\newblock
\showISSN{0045-7825}
\href{https://doi.org/10.1016/j.cma.2008.05.004}{doi:\nolinkurl{10.1016/j.cma.2008.05.004}}
\newblock
\shownote{Computational Methods in Optimization Considering Uncertainties}.


\bibitem[Senapati et~al\mbox{.}(2018)]%
        {cancer}
\bibfield{author}{\bibinfo{person}{Sudipta Senapati},
  \bibinfo{person}{Arun~Kumar Mahanta}, \bibinfo{person}{Sunil Kumar}, {and}
  \bibinfo{person}{Pralay Maiti}.} \bibinfo{year}{2018}\natexlab{}.
\newblock \showarticletitle{Controlled drug delivery vehicles for cancer
  treatment and their performance}.
\newblock \bibinfo{journal}{\emph{Signal Transduction and Targeted Therapy}}
  \bibinfo{volume}{3}, \bibinfo{number}{1} (\bibinfo{year}{2018}),
  \bibinfo{pages}{7}.
\newblock


\bibitem[Sethian(1999)]%
        {sethian1999level}
\bibfield{author}{\bibinfo{person}{James~A Sethian}.}
  \bibinfo{year}{1999}\natexlab{}.
\newblock \bibinfo{booktitle}{\emph{Level Set Methods and Fast Marching
  Methods: Evolving Interfaces in Computational Geometry, Fluid Mechanics,
  Computer Vision, and Materials Science}}.
\newblock \bibinfo{publisher}{Cambridge University Press},
  \bibinfo{address}{Cambridge, UK}.
\newblock
\showISBNx{9780521645577}
\showLCCN{98040859}


\bibitem[Shrishailappa et~al\mbox{.}(2018)]%
        {asthma}
\bibfield{author}{\bibinfo{person}{Doddannavar~Deepika Shrishailappa},
  \bibinfo{person}{Nikhil Tikare}, \bibinfo{person}{Prabhu Halakatti},
  \bibinfo{person}{Anita Desai}, {and} \bibinfo{person}{Iranna Muchandi}.}
  \bibinfo{year}{2018}\natexlab{}.
\newblock \showarticletitle{Formulation and Evaluation of Pulsatile Drug
  Delivery System of Montelukast Sodium}.
\newblock \bibinfo{journal}{\emph{RGUHS Journal of Pharmaceutical Sciences}}
  \bibinfo{volume}{8}, \bibinfo{number}{2} (\bibinfo{year}{2018}),
  \bibinfo{pages}{225--229}.
\newblock


\bibitem[Siepmann and Siepmann(2008)]%
        {drug-delivery}
\bibfield{author}{\bibinfo{person}{Juergen Siepmann} {and}
  \bibinfo{person}{Florence Siepmann}.} \bibinfo{year}{2008}\natexlab{}.
\newblock \showarticletitle{Mathematical modeling of drug delivery}.
\newblock \bibinfo{journal}{\emph{International Journal of Pharmaceutics}}
  \bibinfo{volume}{364}, \bibinfo{number}{2} (\bibinfo{year}{2008}),
  \bibinfo{pages}{328--343}.
\newblock


\bibitem[Siepmann and Siepmann(2012)]%
        {diffusion}
\bibfield{author}{\bibinfo{person}{Juergen Siepmann} {and}
  \bibinfo{person}{Florence Siepmann}.} \bibinfo{year}{2012}\natexlab{}.
\newblock \showarticletitle{Modeling of diffusion controlled drug delivery}.
\newblock \bibinfo{journal}{\emph{Journal of Controlled Release}}
  \bibinfo{volume}{161}, \bibinfo{number}{2} (\bibinfo{year}{2012}),
  \bibinfo{pages}{351--362}.
\newblock


\bibitem[Sigmund(2007)]%
        {regularization2}
\bibfield{author}{\bibinfo{person}{Ole Sigmund}.}
  \bibinfo{year}{2007}\natexlab{}.
\newblock \showarticletitle{Morphology-based black and white filters for
  topology optimization}.
\newblock \bibinfo{journal}{\emph{Structural and Multidisciplinary
  Optimization}} \bibinfo{volume}{33}, \bibinfo{number}{4}
  (\bibinfo{year}{2007}), \bibinfo{pages}{401--424}.
\newblock


\bibitem[Sigmund(2009a)]%
        {sigmund2009manufacturing}
\bibfield{author}{\bibinfo{person}{Ole Sigmund}.}
  \bibinfo{year}{2009}\natexlab{a}.
\newblock \showarticletitle{Manufacturing tolerant topology optimization}.
\newblock \bibinfo{journal}{\emph{Acta Mechanica Sinica}}  \bibinfo{volume}{25}
  (\bibinfo{year}{2009}), \bibinfo{pages}{227--239}.
\newblock


\bibitem[Sigmund(2009b)]%
        {tolerantdesign}
\bibfield{author}{\bibinfo{person}{Ole Sigmund}.}
  \bibinfo{year}{2009}\natexlab{b}.
\newblock \showarticletitle{Manufacturing tolerant topology optimization}.
\newblock \bibinfo{journal}{\emph{Acta Mechanica Sinica - ACTA MECH SINICA}}
  \bibinfo{volume}{25} (\bibinfo{date}{04} \bibinfo{year}{2009}),
  \bibinfo{pages}{227--239}.
\newblock
\href{https://doi.org/10.1007/s10409-009-0240-z}{doi:\nolinkurl{10.1007/s10409-009-0240-z}}


\bibitem[Sigmund and Maute(2013)]%
        {sigmund2013topology}
\bibfield{author}{\bibinfo{person}{Ole Sigmund} {and} \bibinfo{person}{Kurt
  Maute}.} \bibinfo{year}{2013}\natexlab{}.
\newblock \showarticletitle{Topology optimization approaches: A comparative
  review}.
\newblock \bibinfo{journal}{\emph{Structural and multidisciplinary
  optimization}} \bibinfo{volume}{48}, \bibinfo{number}{6}
  (\bibinfo{year}{2013}), \bibinfo{pages}{1031--1055}.
\newblock


\bibitem[Sj{\"o}holm et~al\mbox{.}(2022)]%
        {semi}
\bibfield{author}{\bibinfo{person}{Erica Sj{\"o}holm}, \bibinfo{person}{Rathna
  Mathiyalagan}, \bibinfo{person}{Lisa Lindfors}, \bibinfo{person}{Xiaoju
  Wang}, \bibinfo{person}{Samuli Ojala}, {and} \bibinfo{person}{Niklas
  Sandler}.} \bibinfo{year}{2022}\natexlab{}.
\newblock \showarticletitle{Semi-solid extrusion 3D printing of tailored ChewTs
  for veterinary use - A focus on spectrophotometric quantification of
  gabapentin}.
\newblock \bibinfo{journal}{\emph{European Journal of Pharmaceutical Sciences}}
   \bibinfo{volume}{174} (\bibinfo{year}{2022}), \bibinfo{pages}{106190}.
\newblock


\bibitem[Soprani et~al\mbox{.}(2016)]%
        {mechanical1}
\bibfield{author}{\bibinfo{person}{Stefano Soprani}, \bibinfo{person}{Jan
  Hendrik~Klaas Haertel}, \bibinfo{person}{Boyan~Stefanov Lazarov},
  \bibinfo{person}{Ole Sigmund}, {and} \bibinfo{person}{Kurt Engelbrecht}.}
  \bibinfo{year}{2016}\natexlab{}.
\newblock \showarticletitle{A design approach for integrating thermoelectric
  devices using topology optimization}.
\newblock \bibinfo{journal}{\emph{Applied Energy}}  \bibinfo{volume}{176}
  (\bibinfo{year}{2016}), \bibinfo{pages}{49--64}.
\newblock


\bibitem[Stiepel et~al\mbox{.}(2022)]%
        {erosion}
\bibfield{author}{\bibinfo{person}{Rebeca~T Stiepel}, \bibinfo{person}{Erik~S
  Pena}, \bibinfo{person}{Stephen~A Ehrenzeller}, \bibinfo{person}{Matthew~D
  Gallovic}, \bibinfo{person}{Liubov~M Lifshits},
  \bibinfo{person}{Christopher~J Genito}, \bibinfo{person}{Eric~M Bachelder},
  {and} \bibinfo{person}{Kristy~M Ainslie}.} \bibinfo{year}{2022}\natexlab{}.
\newblock \showarticletitle{A predictive mechanistic model of drug release from
  surface eroding polymeric nanoparticles}.
\newblock \bibinfo{journal}{\emph{Journal of Controlled Release}}
  \bibinfo{volume}{351} (\bibinfo{year}{2022}), \bibinfo{pages}{883--895}.
\newblock


\bibitem[Takahashi and Batty(2022)]%
        {takahashi2022fast}
\bibfield{author}{\bibinfo{person}{Tetsuya Takahashi} {and}
  \bibinfo{person}{Christopher Batty}.} \bibinfo{year}{2022}\natexlab{}.
\newblock \showarticletitle{Fast Marching-Cubes-Style Volume Evaluation for
  Level Set Surfaces}.
\newblock \bibinfo{journal}{\emph{Journal of Computer Graphics Techniques}}
  \bibinfo{volume}{11}, \bibinfo{number}{2} (\bibinfo{year}{2022}),
  \bibinfo{pages}{30--45}.
\newblock


\bibitem[Tan et~al\mbox{.}(2020)]%
        {one-dimentional}
\bibfield{author}{\bibinfo{person}{Yan~JN Tan}, \bibinfo{person}{Wai~P Yong},
  \bibinfo{person}{Jaspreet~S Kochhar}, \bibinfo{person}{Jayant Khanolkar},
  \bibinfo{person}{Xiukai Yao}, \bibinfo{person}{Yajuan Sun},
  \bibinfo{person}{Chi~K Ao}, {and} \bibinfo{person}{Siowling Soh}.}
  \bibinfo{year}{2020}\natexlab{}.
\newblock \showarticletitle{On-demand fully customizable drug tablets via 3D
  printing technology for personalized medicine}.
\newblock \bibinfo{journal}{\emph{Journal of Controlled Release}}
  \bibinfo{volume}{322} (\bibinfo{year}{2020}), \bibinfo{pages}{42--52}.
\newblock


\bibitem[Torres et~al\mbox{.}(2021)]%
        {torres2021robust}
\bibfield{author}{\bibinfo{person}{Alberto~P Torres}, \bibinfo{person}{James~E
  Warner}, \bibinfo{person}{Miguel~A Aguil{\'o}}, {and}
  \bibinfo{person}{James~K Guest}.} \bibinfo{year}{2021}\natexlab{}.
\newblock \showarticletitle{Robust topology optimization under loading
  uncertainties via stochastic reduced order models}.
\newblock \bibinfo{journal}{\emph{Internat. J. Numer. Methods Engrg.}}
  \bibinfo{volume}{122}, \bibinfo{number}{20} (\bibinfo{year}{2021}),
  \bibinfo{pages}{5718--5743}.
\newblock


\bibitem[{U.S. Department of Health and Human Services, Food and Drug
  Administration Center for Drug Evaluation and Research (CDER)}(2015)]%
        {drugsize}
\bibfield{author}{\bibinfo{person}{{U.S. Department of Health and Human
  Services, Food and Drug Administration Center for Drug Evaluation and
  Research (CDER)}}.} \bibinfo{year}{2015}\natexlab{}.
\newblock \bibinfo{title}{Size, Shape, and Other Physical Attributes of Generic
  Tablets and Capsules}.
\newblock


\bibitem[Vermaak et~al\mbox{.}(2014)]%
        {allairelevelset}
\bibfield{author}{\bibinfo{person}{Natasha Vermaak}, \bibinfo{person}{Georgios
  Michailidis}, \bibinfo{person}{Guillaume Parry}, \bibinfo{person}{Rafael
  Estevez}, \bibinfo{person}{Gr{\'e}goire Allaire}, {and} \bibinfo{person}{Yves
  Br{\'e}chet}.} \bibinfo{year}{2014}\natexlab{}.
\newblock \showarticletitle{Material interface effects on the topology
  optimization of multi-phase structures using a level set method}.
\newblock \bibinfo{journal}{\emph{Structural and Multidisciplinary
  Optimization}}  \bibinfo{volume}{50} (\bibinfo{year}{2014}),
  \bibinfo{pages}{623--644}.
\newblock


\bibitem[Wang et~al\mbox{.}(2011)]%
        {smoothHeavisidefunction}
\bibfield{author}{\bibinfo{person}{Fengwen Wang},
  \bibinfo{person}{Boyan~Stefanov Lazarov}, {and} \bibinfo{person}{Ole
  Sigmund}.} \bibinfo{year}{2011}\natexlab{}.
\newblock \showarticletitle{On projection methods, convergence and robust
  formulations in topology optimization}.
\newblock \bibinfo{journal}{\emph{Structural and Multidisciplinary
  Optimization}} \bibinfo{volume}{43}, \bibinfo{number}{6}
  (\bibinfo{year}{2011}), \bibinfo{pages}{767--784}.
\newblock


\bibitem[Wang and Wang(2005)]%
        {regularization}
\bibfield{author}{\bibinfo{person}{Michael~Yu Wang} {and}
  \bibinfo{person}{Shengyin Wang}.} \bibinfo{year}{2005}\natexlab{}.
\newblock \showarticletitle{Bilateral filtering for structural topology
  optimization}.
\newblock \bibinfo{journal}{\emph{Internat. J. Numer. Methods Engrg.}}
  \bibinfo{volume}{63}, \bibinfo{number}{13} (\bibinfo{year}{2005}),
  \bibinfo{pages}{1911--1938}.
\newblock


\bibitem[Warner et~al\mbox{.}(2013)]%
        {sromgrigoriu2}
\bibfield{author}{\bibinfo{person}{James~E. Warner}, \bibinfo{person}{Mircea
  Grigoriu}, {and} \bibinfo{person}{Wilkins Aquino}.}
  \bibinfo{year}{2013}\natexlab{}.
\newblock \showarticletitle{Stochastic reduced order models for random vectors:
  Application to random eigenvalue problems}.
\newblock \bibinfo{journal}{\emph{Probabilistic Engineering Mechanics}}
  \bibinfo{volume}{31} (\bibinfo{year}{2013}), \bibinfo{pages}{1--11}.
\newblock
\showISSN{0266-8920}
\href{https://doi.org/10.1016/j.probengmech.2012.07.001}{doi:\nolinkurl{10.1016/j.probengmech.2012.07.001}}


\bibitem[Zhu et~al\mbox{.}(1997)]%
        {optimizer}
\bibfield{author}{\bibinfo{person}{Ciyou Zhu}, \bibinfo{person}{Richard~H
  Byrd}, \bibinfo{person}{Peihuang Lu}, {and} \bibinfo{person}{Jorge Nocedal}.}
  \bibinfo{year}{1997}\natexlab{}.
\newblock \showarticletitle{Algorithm 778: L-BFGS-B: Fortran subroutines for
  large-scale bound-constrained optimization}.
\newblock \bibinfo{journal}{\emph{ACM Transactions on Mathematical Software
  (TOMS)}} \bibinfo{volume}{23}, \bibinfo{number}{4} (\bibinfo{year}{1997}),
  \bibinfo{pages}{550--560}.
\newblock


\bibitem[Zunino et~al\mbox{.}(2025)]%
        {eikonalinfo3}
\bibfield{author}{\bibinfo{person}{Andrea Zunino}, \bibinfo{person}{Scott
  Keating}, {and} \bibinfo{person}{Andreas Fichtner}.}
  \bibinfo{year}{2025}\natexlab{}.
\newblock \bibinfo{title}{A discrete adjoint method for deterministic and
  probabilistic eikonal-equation-based inversion of travel time for velocity
  and source location}.
\newblock
\showeprint[arxiv]{2501.13532}


\end{thebibliography}

\end{document}